\documentclass[aip, twocolumn, reprint, amsmath, superscriptaddress,showkeys,showpacs]{revtex4-1}
 
 \usepackage[colorlinks=true,
 linkcolor=red,
 urlcolor=black,
 citecolor=blue]{hyperref}
 \usepackage{graphicx}
 \usepackage{epstopdf}
 \usepackage{dcolumn}
 \usepackage{bm, soul}
 \usepackage{caption}
 \usepackage{subcaption}
 \usepackage{dsfont}
 \usepackage[toc]{appendix}
 \usepackage{ragged2e}
 \usepackage{float}
 \DeclareCaptionJustification{justified}{\justifying}
\begin{document}
	
	\title{Routes to extreme events in dynamical systems: Dynamical and Statistical Characteristics}
	\author{Arindam Mishra}
    \affiliation{Department of Mathematics, Jadavpur University, Jadavpur,  Kolkata  700032, India}
	\author{S Leo Kingston}
	\affiliation{Division of Dynamics, Lodz University of  Technology, 90-924 Lodz, Poland}
	\author{Chittaranjan Hens}
	\affiliation{Physics and Applied Mathematics Unit, Indian Statistical Institute, Kolkata 700108, India}
	\author{Tomasz Kapitaniak}
	\affiliation{Division of Dynamics, Lodz University of  Technology, 90-924 Lodz, Poland}
	\author{Ulrike Feudel}
	\affiliation{Institute for Chemistry and Biology of the Marine Environment, University of Oldenburg, 
26111 Oldenburg, Germany}
	\author{Syamal K. Dana}
\affiliation{Division of Dynamics, Lodz University of  Technology, 90-924 Lodz, Poland}
	\affiliation{Department of Mathematics, Jadavpur University, Jadavpur,  Kolkata  700032, India}
		
	\date{\today}
	\begin{abstract}
   Intermittent large amplitude events are seen in the temporal evolution of a state variable of many dynamical systems. Such intermittent large events suddenly start appearing in dynamical systems at a critical value of a system parameter and continues for a range of parameter values. Three important processes of instabilities, namely, interior crisis, Pomeau-Manneville intermittency and the breakdown of quasiperiodic motion, are most common as observed in many systems that lead to such occasional and rare transitions to large amplitude spiking events. We characterize these occasional large events as extreme events if they are larger than a statistically defined significant height. We present two exemplary systems, a single system and a coupled system to illustrate how the instabilites work to originate as extreme events and they manifest as non-trivial dynamical events. We illustrate the dynamical and statistical properties of such events.	\end{abstract}
	
	\maketitle
\begin{quotation}
Extreme events such as floods, tsunamis, cyclones, power-grid failures, share market crashes are very well known  natural disasters that have been  addressed by different groups, governmental and societal, in search of a solution how to mitigate them for  saving life and economy.  These events are rare, recurrent and extraordinarily large in  size. Such disastrous events also occur in humans as epileptic seizures in the brain. Attempts have been made  to develop a basic understanding of the dynamics of such  natural phenomena using simple models and  to search for possible techniques for their prediction. In this endeavor, extreme events  are seen as a sudden and intermittent rising of a systems' amplitude to large values from a nominally low amplitude state. From this perspective, we recognize that extreme event-like behavior can emerge in dynamical systems via three  well known generic routes to the origin of instability in dynamical systems, which were not always focused as exemplary mechanisms of the origin of extreme events. We emphasize here  that these three common routes, namely, interior-crisis-induced, intermittency-induced and quasiperiodic-breakdown may lead to extreme events in many dynamical systems and we elaborate the mechanisms here using two model examples.
\end{quotation}

\section{Introduction}

\par Understanding the emergence and characteristics of extreme events is a pressing problem in different disciplines of science because of their possible large impact; well-known examples are earthquakes, floods, tsunamis, cyclones \cite{easterling},  share market crashes \cite{kantz}. 
Such extremely large devastating events are rare, but recurrent. Usually extreme events possess  a size larger than 4-8 times the standard deviation of the nominal size of events. However, even smaller size events such as the epileptic seizures in the human brain \cite{lehnertz, Klaus}, can have a devastating effect, but then other measures need to be adopted to detect the extreme behavior in the brain. Basically extremes depict a situation of significant departure from a nominal state of a system that causes a disaster. In statistics, extreme events are often defined as the set of events being larger than the 90-99$^{th}$ percentile of their probability distribution \cite{mcphillips}. The discussion about a possible increase in extreme events like hurricanes, strong rainfall events \cite{trenberth,goswami} or rogue waves \cite{Akhmediev}, in the course of climate change encouraged many researchers to explore the  mechanisms of the emergence of extreme event-like dynamical behaviors in model systems \cite{kharif,muller,krause,biggs,Akhmediev,dudley,ott,ansmann,messori,farazmand2,mulhern} to develop a better understanding of the phenomenon from a dynamical systems point of view and for developing possible methods of their prediction \cite{sapsis1, karnatak2}. Laboratory-scale experiments were also set-up using electronic circuits \cite{kingston}, optical systems \cite{bonatto,cristina,reinoso, solli,Tlidi,ref7,clerc}, experiments in a wave channel \cite{Hoffmann} to create such occasional large amplitude events to study in detail their dynamic origin. Multistable systems possessing a multitude of coexisting attractors  show  occasional large amplitude events when the trajectory of the dynamical system wanders \cite{pisarchik} between coexisting stable periodic attractors of varying amplitude under the influence of noise. Another class of extreme events emerges in systems with discontinuous boundaries that exhibit stick-slip dynamics \cite{kumarasamy}. Extreme event-like behaviors have also been demonstrated in coupled systems \cite{ott,karnatak,doedel} and networks of dynamical systems \cite{rogister,ansmann,mishra,hens}. Recently, strategies for controlling or even suppressing emergence of extreme events have been attempted  \cite{suresh,kishore,farazmand1,galuzio, ott, sorbendu}.
\par The most important question is how such extremely large and rare events emerge in dynamical systems? By this time, it has been established that the basic mechanism \cite{sapsis, sapsis1} of the origin of extreme events in dynamical systems, in general, lies in the existence of a region of instability in state space of a system, which a chaotic trajectory may occasionally visit resulting in travels to locations in state space far away from the original attractor to which the trajectory returns after a short duration.  This  is  manifested as occasional large amplitude events in the time evolution of systems, which have an analogy to sudden and rare large events in nature. 
A more fundamental question then arises: What are the types of instabilities that lead to such occasional extremely large events in dynamical systems? In fact, we learn from the recent works that the types of instabilities are manifold. Firstly, a type of interior crisis has been identified in many recent works \cite{cristina, kingston, karnatak, bonatto} as a cause of distant excursions leading to occasional large events. Secondly, a loss of transverse stability of certain periodic orbits in the synchronization manifold of coupled systems can lead to the formation of extreme events \cite{saha1,blackbeard}. This transition is closely related to the bubbling transition \cite{venkatarami,ashwin}. Thirdly, extreme deviations from the normal behavior, which can also be considered as an extreme event, can occur in slow-fast systems \cite{sapsis1,wieczorek}. Fourthly, extreme events are found to originate in response to parameter variation, namely, via the traditional  Pomeau-Manneville (PM) intermittency \cite{kingston} and quasiperiodic breakdown \cite{mishra} routes to chaos. In general, all these mechanisms can lead to the formation of extreme events manifested as large rare and recurrent excursions from the normal behavior. However, any apparently large event occurring in a system does not always qualify to be called  an extreme event because the size of the excursion might not be always so large in amplitude to satisfy a statistical condition. Our focus here is on systems with extremely large events only. In such cases, the amplitude has to be larger than 4-8 times the standard deviation from the nominal size of events or it has to be found in the 90-99th percentile of the probability distribution. Whether this statistical criterion is met depends on the system under consideration. 
\par The interior crisis is manifested as a sudden enlargement of the size of a chaotic attractor when a variation of a system parameter leads to crossing a critical threshold \cite{celso}. This type of crisis is due to a collision of a chaotic attractor with the stable manifold of an unstable fixed point or an unstable periodic orbit. It was first illustrated in the Ikeda map \cite{celso}, where it was shown  that the chaotic trajectory of the system remains bounded to a certain region in state space for a range of system parameters, but starts occasional excursions to far away locations when the parameter is crossing the critical value.
This temporal dynamics is reflected in a chaotic oscillation of smaller amplitude, but with intermittent large amplitude spikes. These  occasional large amplitude spikes in the Ikeda map exhibit an analogy to the occurrence of extreme events \cite{sorbendu}. 
A different kind of interior crisis originates from a collision of a period-adding and a period doubling cascade that has been studied in diffusively coupled \cite{karnatak} as well as delay-coupled \cite{saha} relaxation oscillators such as  FitzHugh-Nagumo oscillator. A similar mechanism related to the existence of a homoclinic orbit and intermittent large events has also been investigated in a climate model \cite{arnob1}. While crisis-induced extreme events have widely been discussed in many examples of current studies \cite{karnatak, cristina, bonatto,kingston}, PM intermittency-induced sudden expansion of attractor has been ignored \cite{reinoso}, although indication of such a possible route to extreme events has been clearly present in a laser system given a closer inspection in the system's response to parameter variation in a bifurcation diagram \cite{reinoso}. PM intermittency \cite{PM} is a well understood phenomenon in dynamical systems, in models as well as in experimental and natural systems, which shows intermittent laminar and turbulent phases in the time evolution. The examples of systems with PM intermittency are abundant in literature; some of them really showed intermittent very large events, but none of them were really looked into from the angle of extreme events except recently in experiments of thermo-acoustic systems \cite{sujith}. 
\par   Here we present and compare different  possible types of instabilities leading to extreme events in dynamical systems to find commonalities and differences between them. We find  interior crisis-induced extreme events as the most common route in many dynamical systems, but also find two additional routes, PM intermittency and breakdown of quasiperiodicity that also lead to extreme events. These last two routes  were not discussed, in the literature, in the context of their role in extreme events. We find a mention of these three routes, in the earlier works of Nicolis et al \cite{nicolis} in connection with the studies of extreme events using the statistical properties in maps, but they have not elaborated the specific dynamical properties leading to the formation of extreme events. We elaborate here the dynamical and statistical properties of all the three routes by employing two different systems, each of them exhibiting two different routes to extreme events providing sufficient dynamical evidence of the routes they \cite{nicolis} predict. 
Firstly, we elaborate the crisis-induced and the intermittency-induced routes to extreme events in a single Li\'enard system \cite{kingston}. Secondly, we present the example of two coupled Hindmarsh-Rose (HR) neuron models \cite{HR} under mutual synaptic interactions \cite{mishra}. Here  we find evidence of the PM intermittency route to extreme events once again as well as the third kind of instability that leads to extreme events via the breakdown of quasiperiodic motion. This last route has not been explored so far from the dynamical point of view of extreme events, to the best of our knowledge.
 \par In summary, we make an attempt to answer the basic question what are the possible types of instabilities that originate in  dynamical systems in response to parameter variation, leading to extreme events? We discuss here three most common types of instabilities: interior crisis, PM intermittency and break-down of quasiperiodicity that  can lead to the origin of extreme events, in many systems.  These three routes are illustrated with two example systems. All three kinds of instabilities described 
here are generic in nature but will lead to extreme events only in specific dynamical systems. The reason for this behavior lies in the fact, that the excursions in state space which result from those instabilities are not always that large that they would satisfy the statistical conditions on their magnitude. Whether an event occurring is extreme 
depends on the system under consideration and its intrinsic dynamical traits.
 	\begin{figure}
 	\includegraphics[height=7.5cm, width=8.75cm]{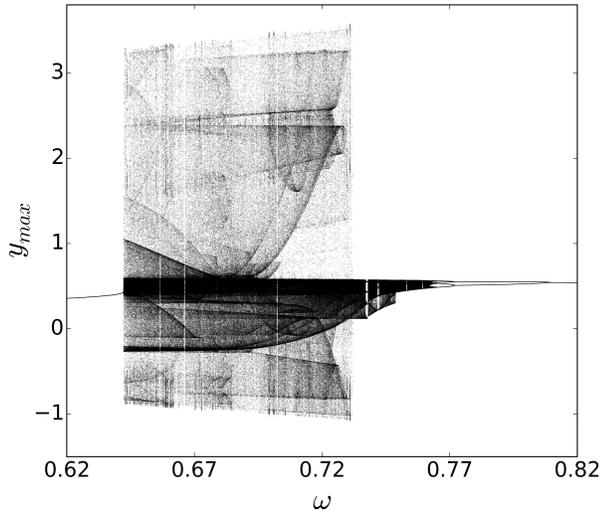} 
 	\caption{(color online) 
 		Bifurcation diagram of $y_{max}$ against forcing frequency $\omega$. A period-doubling cascade to chaos for decreasing frequency is followed by a sudden explosion at $\omega$ = 0.7315 at right. The expanded amplitude continues, but  drops to low amplitude period-1 oscillation at $\omega$=0.6423 shown at left. 
 	}
 \label{FIG.1}
 \end{figure}
   \section{Extreme Events: interior crisis route}
 We first elaborate the case of the period-doubling route to chaos followed by an interior crisis leading to extreme events in a forced Li\'enard type system,  
 \begin{gather*} 
 \dot{x}=y, \nonumber \\ 
 \dot{y}=-\alpha{x}y-\gamma{x}-\beta{x^3}+A sin(\omega t),
 \end{gather*} 
 $\alpha$ and $\beta$ are  nonlinear damping and  strength of nonlinearity, respectively, $\gamma$ is related to the intrinsic frequency of the autonomous system, and $A$ is the amplitude and $\omega$ is the frequency of the periodic forcing.   
 The autonomous system shows \cite{kingston, mishra1} a dual character  of dissipative and conservative dynamics for a set of parametric conditions, $\alpha>0$, $\beta>0$ and $\gamma<0$ (more details on the autonomous system's behavior are presented in Appendix A). We arbitrarily select the parameters, $\alpha~\mathrm{= 0.45}$, $\beta~\mathrm{= 0.50}$ and  $\gamma~\mathrm{= -0.50}$ that satisfy  the conditions to maintain the dual character.  The autonomous system has a homoclinic orbit (HO) connecting the saddle origin. Outside this HO there are infinitely large number of neutrally stable orbits. 
 When a periodic forcing is applied to the system, 
each of the neutrally stable
 orbits outside the HO turns into a quasiperiodic orbit lying on top of each other in many layers that coexist with a periodic or chaotic orbit inside the HO depending upon the forcing frequency. The saddle point (0,0) becomes a saddle orbit \cite{guckenheimer}. 
\begin{figure*}[]
		\hspace{-130pt}
		\begin{subfigure}[b]{0.5\columnwidth}
			\centering
			\includegraphics[width = 9cm,height = 6cm]{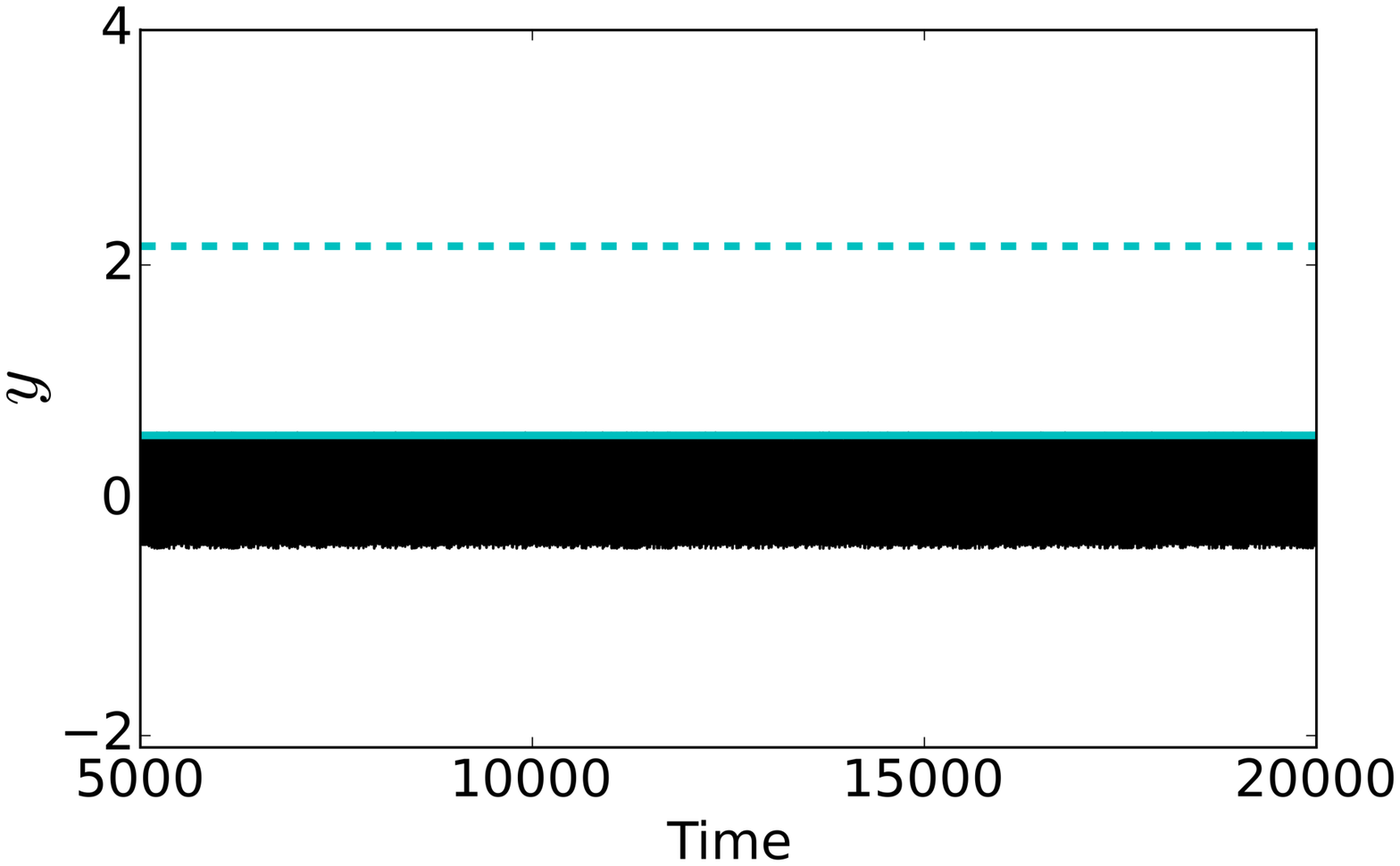}
			\label{}	
		\end{subfigure}
		\hspace{130pt}
		\begin{subfigure}[b]{0.5\columnwidth}
			\centering
			\includegraphics[width = 7cm,height = 6cm]{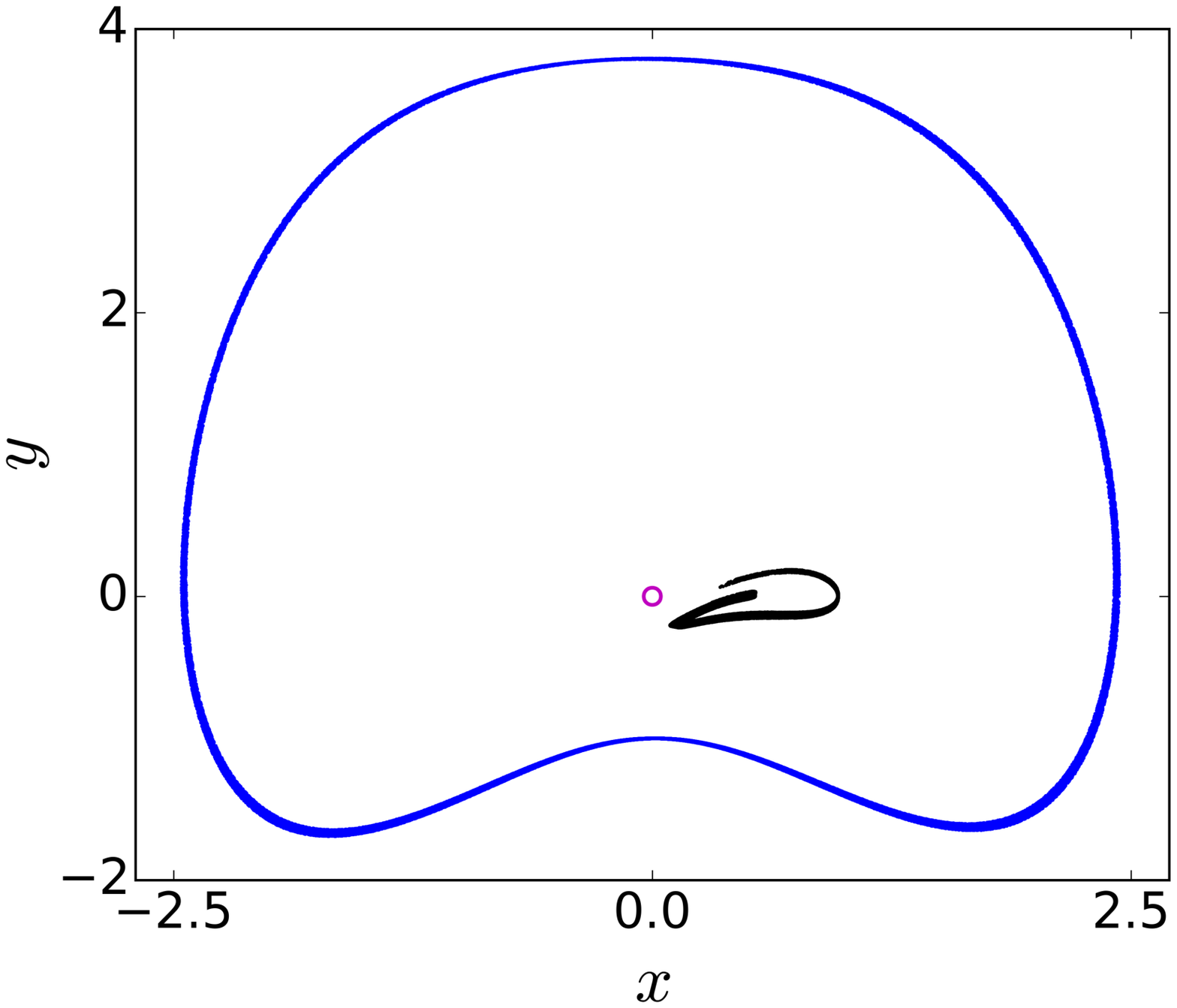}
			\label{}	
		\end{subfigure}\\
	\hspace{-130pt}
	\begin{subfigure}[b]{0.5\columnwidth}
		\centering
		\includegraphics[width = 9cm,height = 6cm]{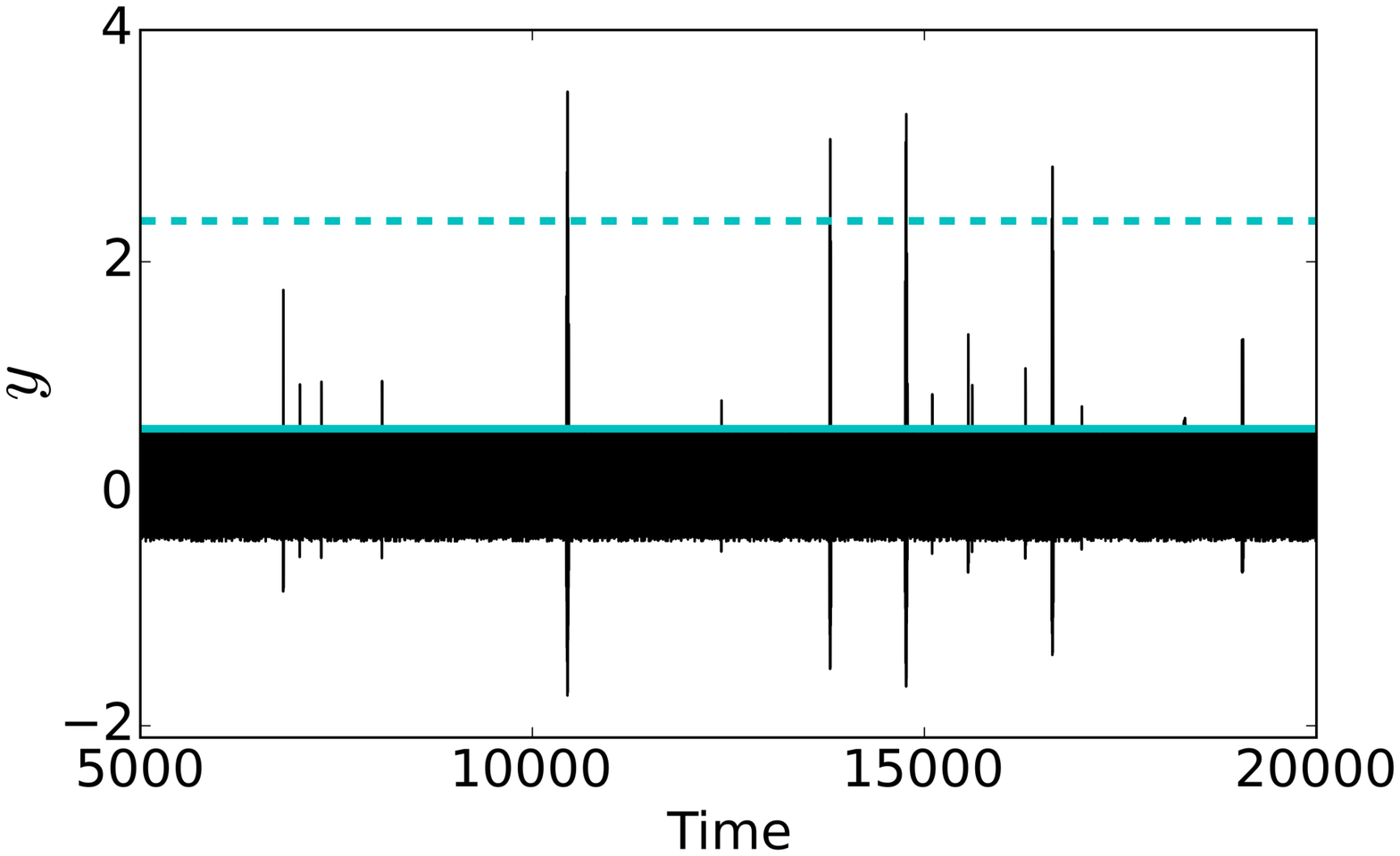}
		\label{}	
	\end{subfigure}
	\hspace{130pt}
	\begin{subfigure}[b]{0.5\columnwidth}
		\centering
		\includegraphics[width = 7cm,height = 6cm]{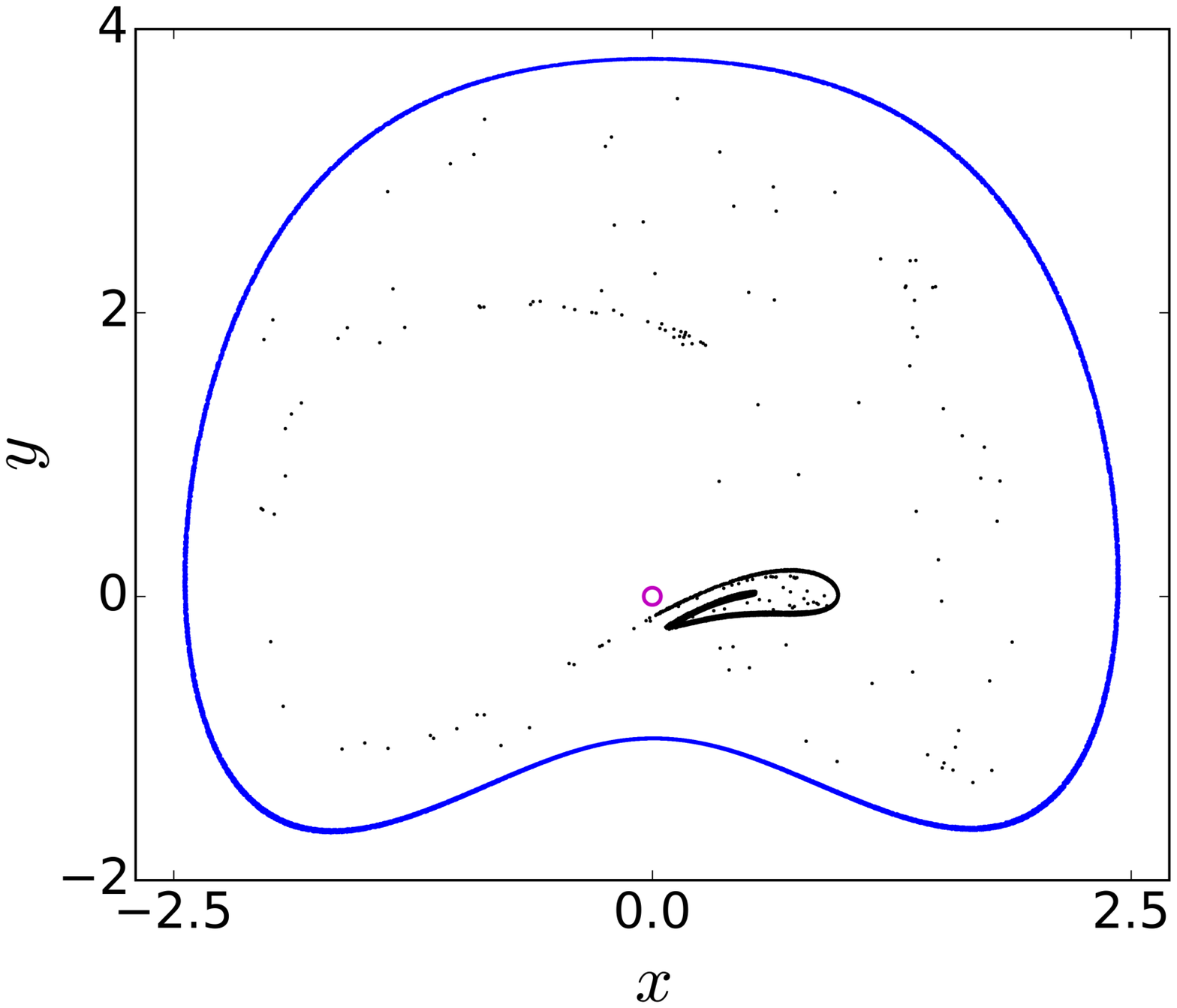}
		\label{}	
	\end{subfigure}\\
\hspace{-130pt}
\begin{subfigure}[b]{0.5\columnwidth}
	\centering
	\includegraphics[width = 9cm,height = 6cm]{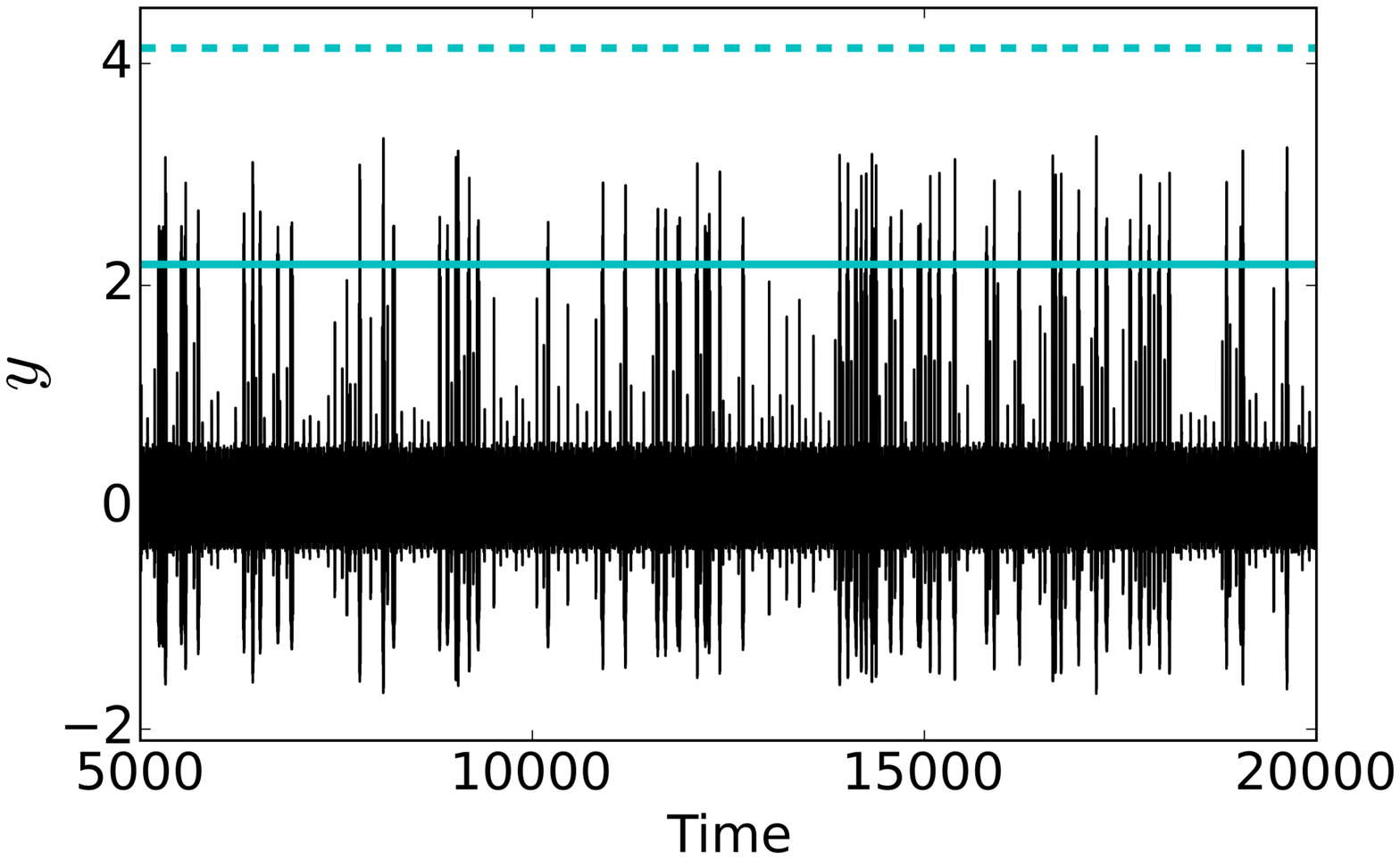}
	\label{}	
\end{subfigure}
\hspace{130pt}
\begin{subfigure}[b]{0.5\columnwidth}
	\centering
	\includegraphics[width = 7cm,height = 6cm]{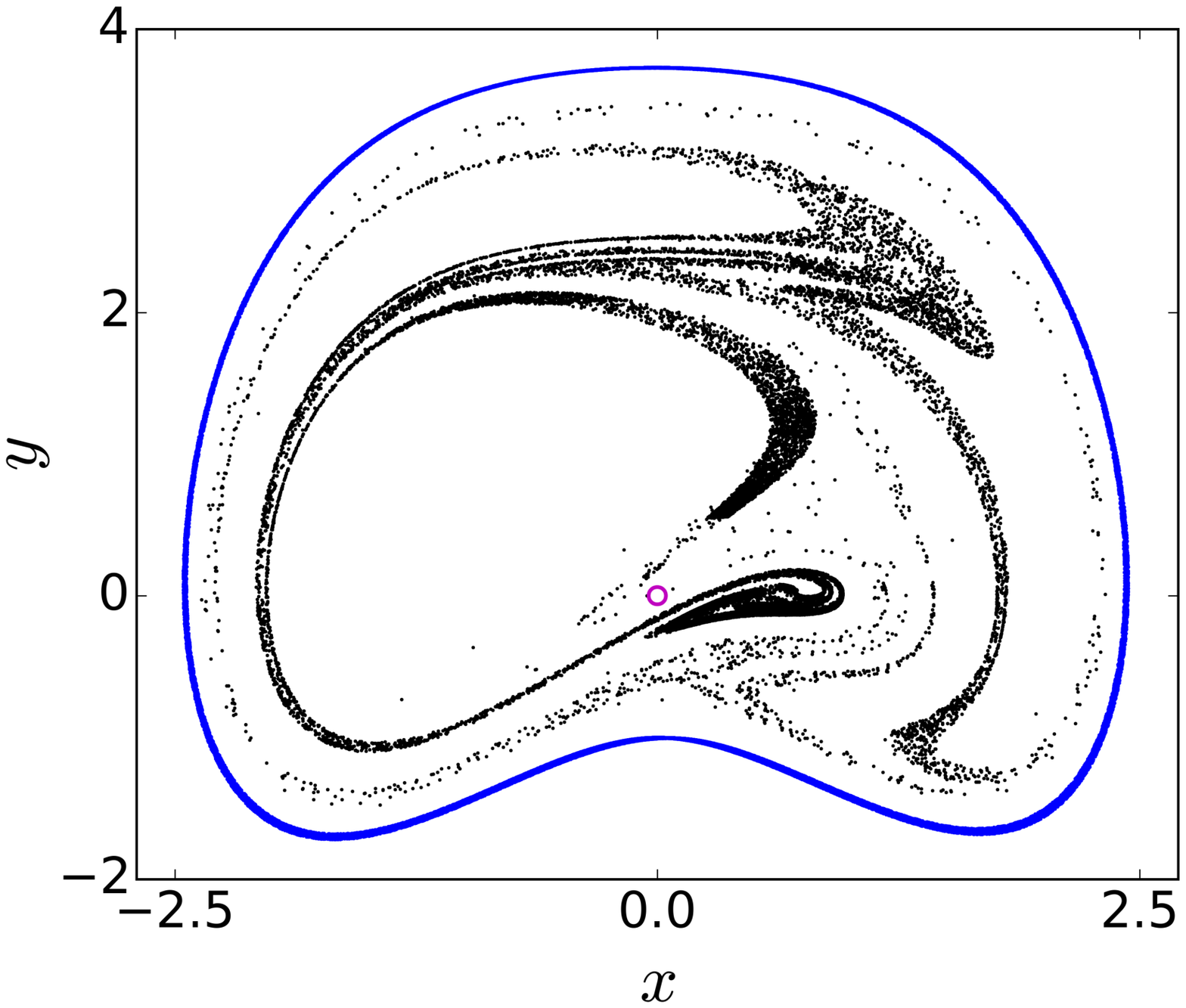}
	\label{}	
\end{subfigure}
\caption{(Color online) Time series (left panels) and Poincar\'e surface of section plot (right panels) of the forced Li\'enard system for the interior crisis route. Quasiperiodic orbits as a closed curve (blue color), chaotic orbits in black dots. Open circles (magenta color) denotes a saddle orbit (right panels). Solid and dashed line in cyan denote 99$^{th}$ percentile and $H_{s}$ line respectively. Upper panels ($\omega = 0.735$) shows bounded chaotic attractor (black dots), middle panels ($\omega = 0.7315$ shows a collision of chaotic orbit with the saddle orbit around origin leading to extreme events indicated by scattered points representing occasional far-away excursion of the trajectory within the quasiperiodic  boundary (curve in blue). Lower panels correspond to a scenario at $\omega = 0.7$ where spiking occurs more frequently (left panel) and there is no extreme event as  $H_{s}$ line is well above. The lower right panel shows densely populated points (black) representing frequent spiking events whose amplitudes are limited within the quasiperiodic boundary (blue curve). }
	\label{FIG.2}
\end{figure*}

\begin{figure*}[]
	\hspace{-130pt}
	\begin{subfigure}[b]{0.5\columnwidth}
		\centering
		\includegraphics[width = 9cm,height = 6cm]{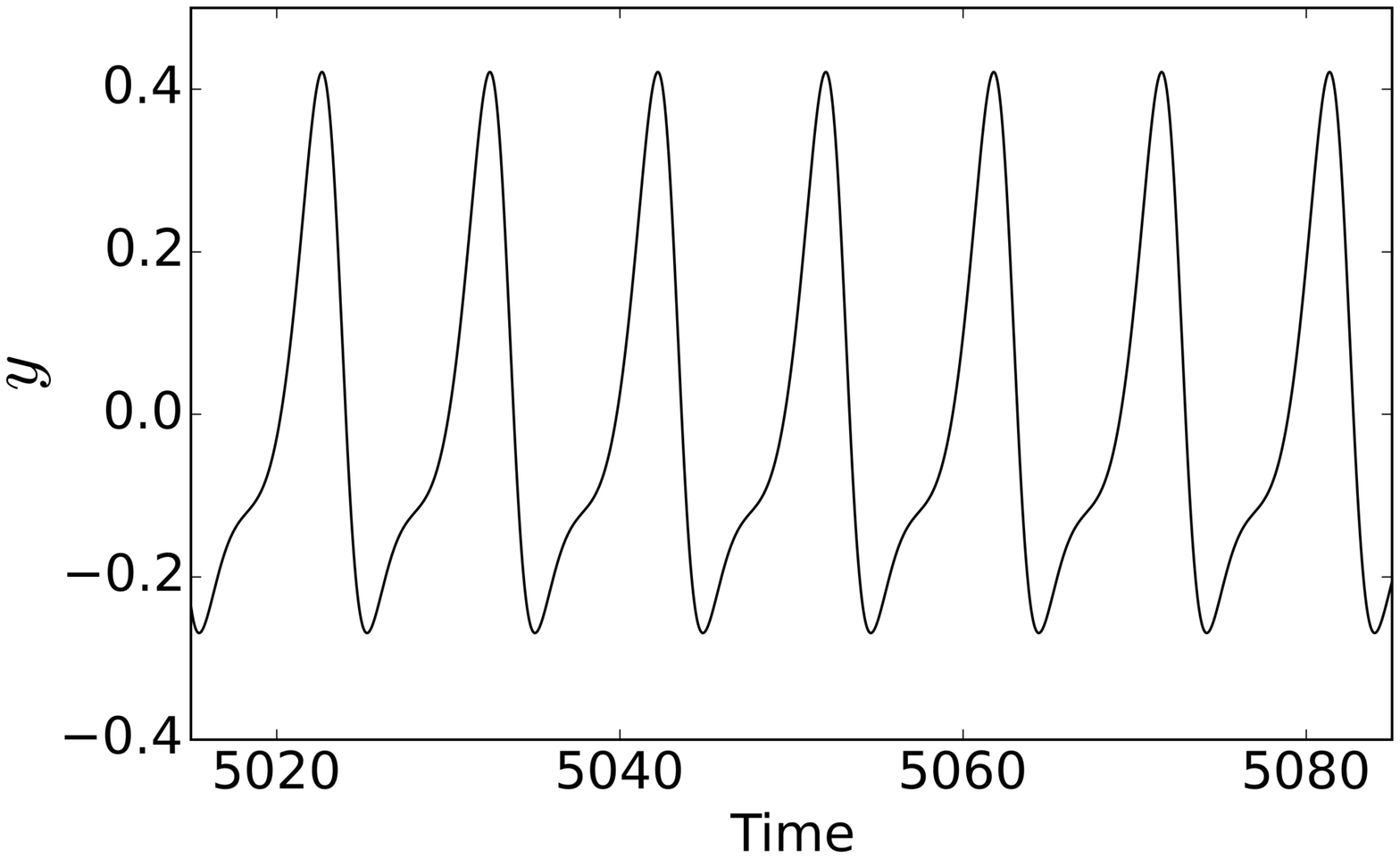}
		\label{}	
	\end{subfigure}
	\hspace{130pt}
	\begin{subfigure}[b]{0.5\columnwidth}
		\centering
		\includegraphics[width = 7cm,height = 6cm]{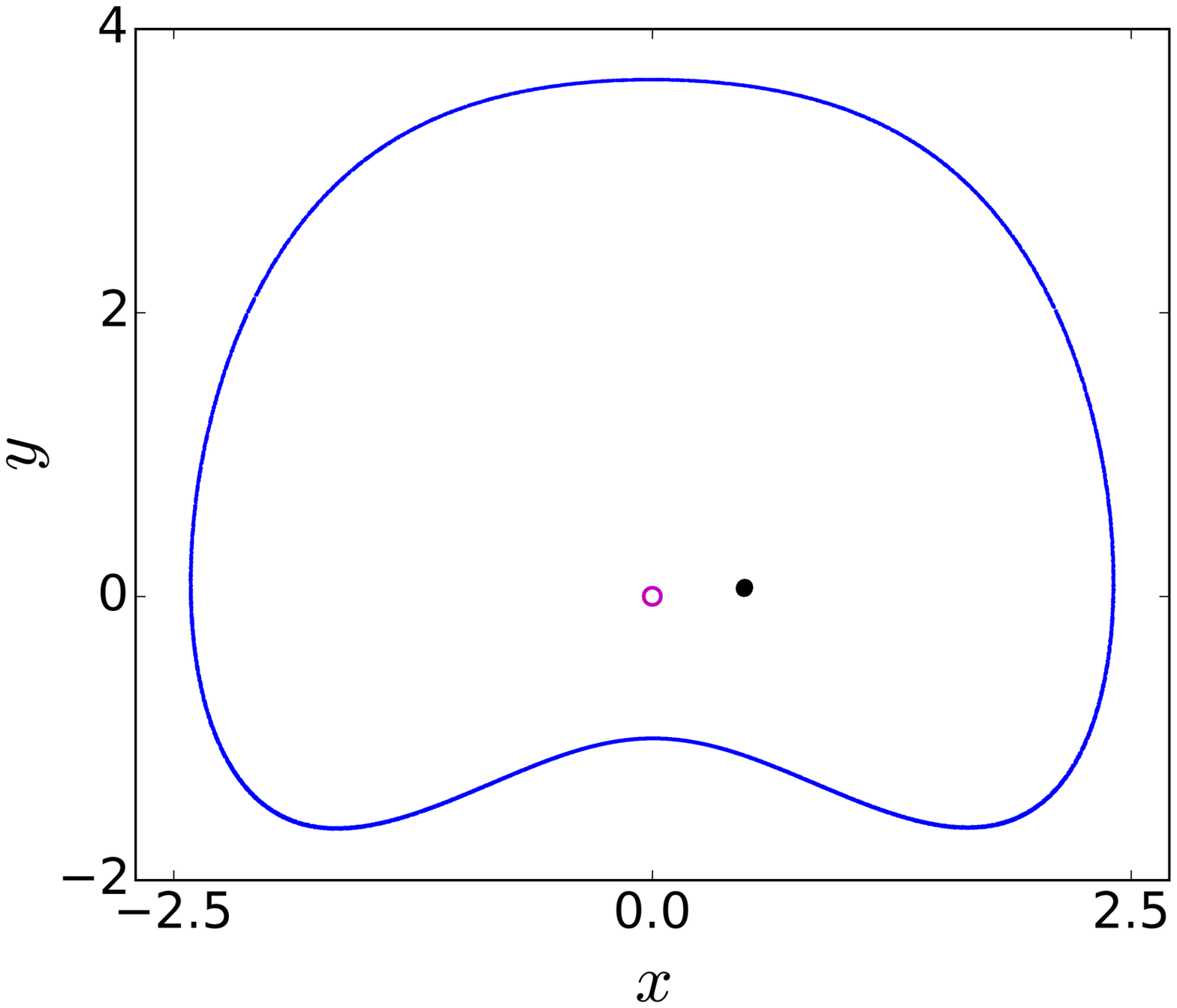}
		\label{}	
	\end{subfigure}\\
	\hspace{-130pt}
	\begin{subfigure}[b]{0.5\columnwidth}
		\centering
		\includegraphics[width = 9cm,height = 6cm]{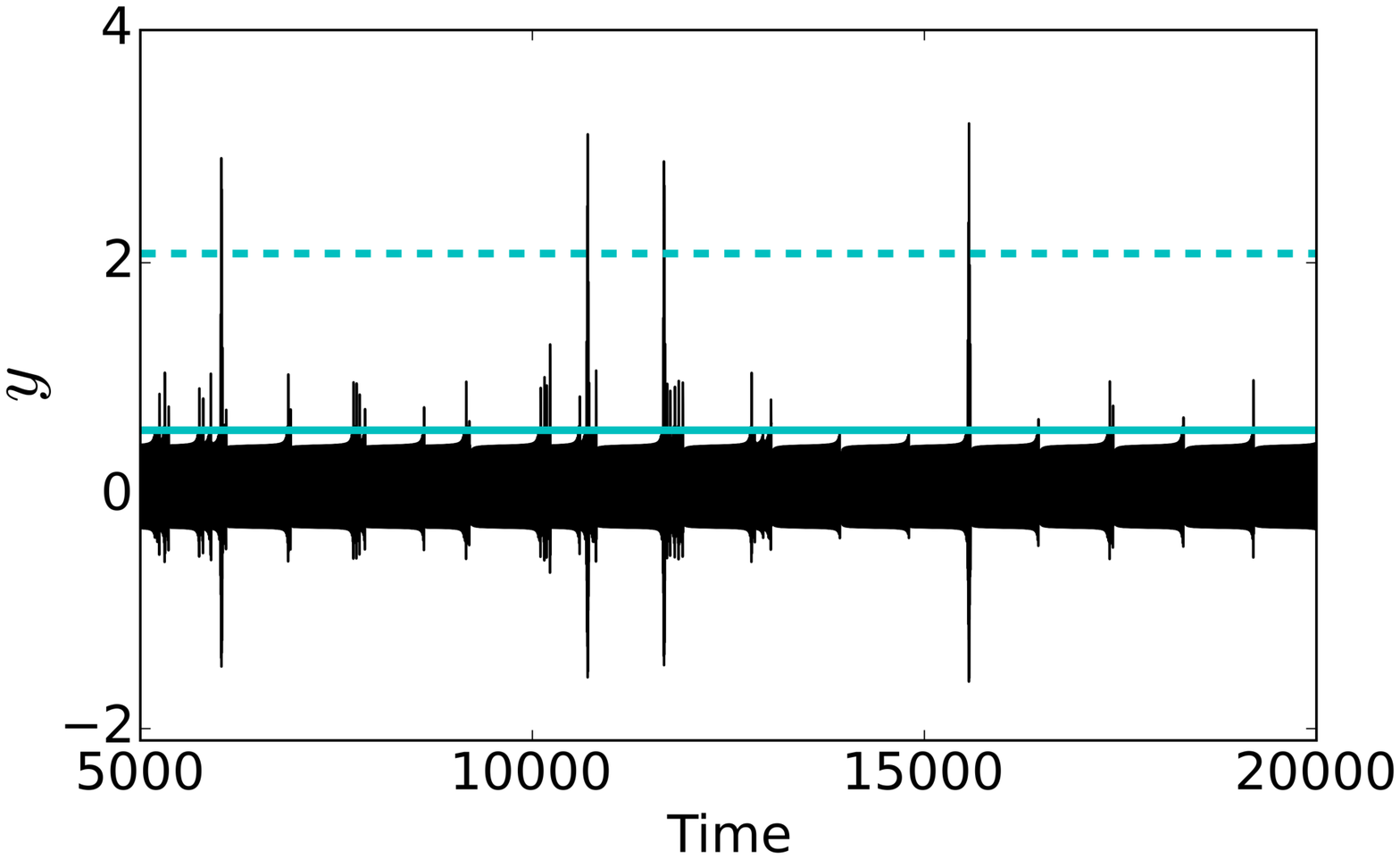}
		\label{}	
	\end{subfigure}
	\hspace{130pt}
	\begin{subfigure}[b]{0.5\columnwidth}
		\centering
		\includegraphics[width = 7cm,height = 6cm]{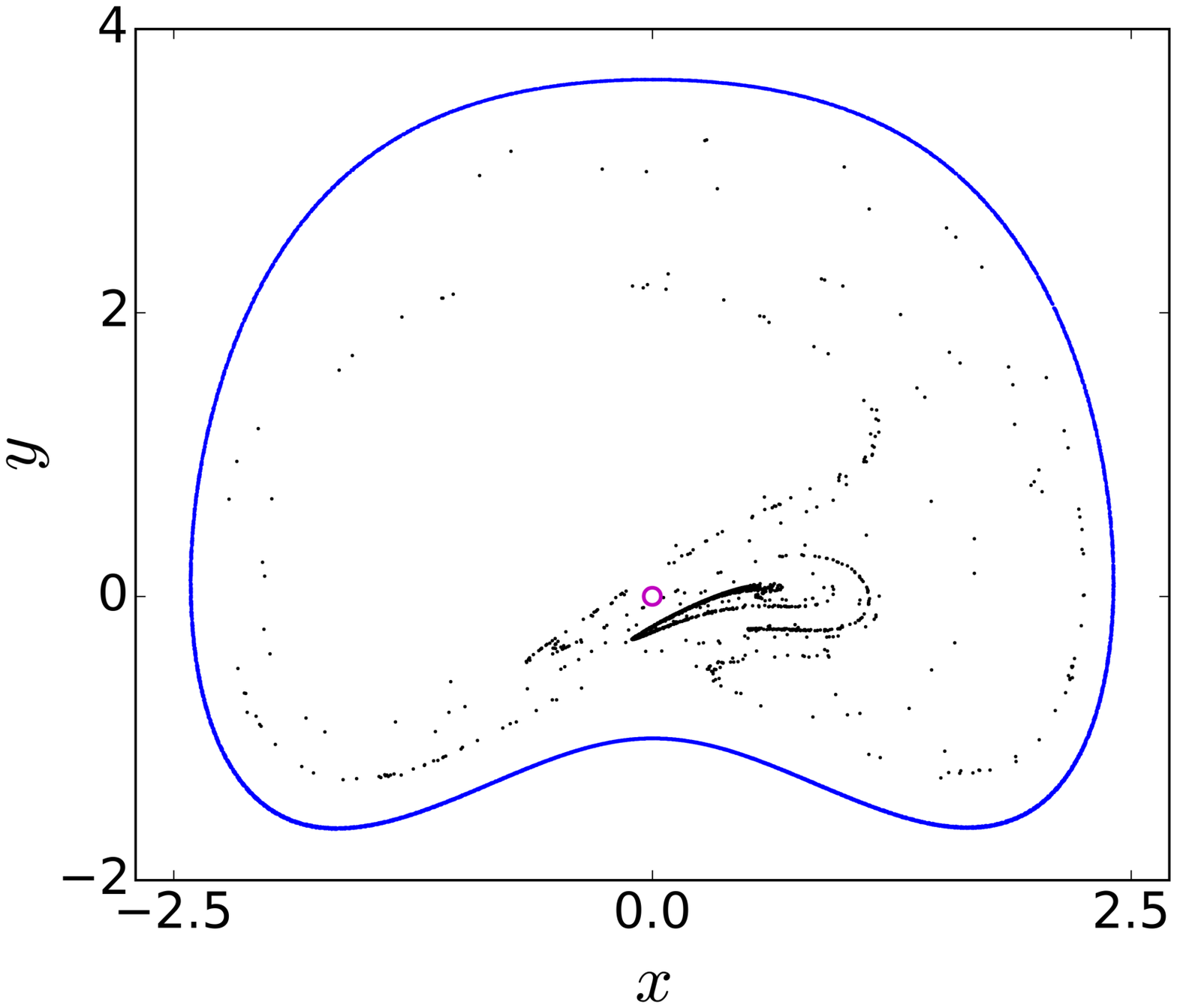}
		\label{}	
	\end{subfigure}
	\caption{(Color online) Time series (left panels) and Poincar\'e surface of sections plot (right panels) of the forced Li\'enard system for PM intermittency. The Poincar\'e plots represent the quasiperiodic motion in closed curve (blue color), chaotic orbits in black dots. Open circles (magenta color) denote saddle orbit. Upper panels ($\omega = 0.642$) shows periodic time series (left panel) and Poincar\'e plot  (right panel) shows quasiperiodic motion (closed blue curve) and saddle orbit(open circle) and periodic orbit (closed circle). The 
		lower panels ($\omega =0.6423$) shows extreme events (time series at left panel) with many scattered points (right panel, black dots) within the quasiperiodic boundary (blue curve).}
	\label{FIG.3}
\end{figure*}
For an appropriate choice of initial conditions, from inside the HO, we produce a bifurcation diagram of the forced system, as shown in Fig.~1, depending on the forcing frequency $\omega$ in a range of values ($0.62$ to $0.82$). The stable focus of the autonomous system at $(1,0)$ first becomes unstable and turns into a stable periodic orbit, which subsequently undergoes a period-doubling cascade into chaos with decreasing $\omega$. The chaotic dynamics (dense black region), although slowly increasing its size with decreasing $\omega$, remains bounded 
until a critical point is reached when $y_{max}$  (maxima of $y$=$y_{max}$), in the bifurcation diagram,  suddenly explodes indicating a very large amplitude of  oscillation at $\omega=0.7315$. 
The expanded attractor continues to exist with decreasing frequency until it reaches another critical point $\omega=0.6423$ at which chaos suddenly transits into a low amplitude periodic oscillation with a sudden large drop in the $y_{max}$ value. 
 Choosing another set of initial conditions outside the HO, we obtain quasiperiodicity  originating from the periodic orbits (neutrally stable orbits) for the whole range of forcing frequency from high to low (not shown here, refer to \cite{kingston}).
 Large and sudden change in $y_{max}$ is seen at two critical forcing frequencies $\omega$=($0.7315$ and $0.6423$) at two ends of the bifurcation diagram.  
 A similar sudden expansion of the attractor in response to parameter variation was reported earlier in a laser system \cite{reinoso} that  shows a period-doubling route to chaos followed by a large expansion of the attractors due to an interior crisis and the origin of extreme events. We shall provide evidence that a similar interior-crisis also plays a key role in the origin of extreme events, in our example systems, at this critical frequency $\omega=0.7315$. On the other hand, when we focus at the critical point $\omega=0.6423$ (left of Fig.~1), we find an intermittency-induced sudden expansion that also leads to the origin of extreme events, which we explain further in the next section. Such a intermittency-induced sudden expansion in the bifurcation diagram can easily be noticed in the earlier study \cite{reinoso}, but it has been ignored by the authors.
\par The bifurcation diagram (Fig.~1) could be misleading as giving an impression of a permanent expansion of the attractor size of the system  at  critical values of $\omega$. A closer inspection of the bifurcation diagram will reveal that the original low amplitude chaotic dynamics continues (densely occupied black region) for the whole range of forcing frequencies, $\omega=(0.7315$ to $0.6423)$, that accompanies large values of $y_{max}$, as represented by a sparse cloud of points, especially, near the edges, which indicate occasional large jump of the attractor size. It is confirmed when we observe the temporal dynamics of the system for frequencies near the edges of the expanded bifurcation diagram in Fig.~1. The bounded chaos with low amplitude  is reflected in the temporal dynamics in Fig.~2 (upper row, left panel) for a  larger $\omega$=0.735 away from the critical value. The intermittent large expansion is confirmed in the temporal dynamics in Fig.~2 (middle row,  left panel) for a critical frequency $\omega=0.7315$ that shows  occasional large spikes emanating from bounded chaos. A horizontal dashed line (cyan color) is drawn that defines a significant height, $H_S$= $\langle P_{n}\rangle$ + $8\sigma$, where  $\langle P_{n}\rangle$ is the average of all the peaks in a very long time series of $y$ and $\sigma$ is the standard deviation. If a spike crosses the $H_S$ line, we call it an extreme event when it occurs recurrently, but rarely. A second solid horizontal line (cyan color)  is drawn to mark the $99^{th}$ percentile line above which statisticians would call the events as extreme. The $99^{th}$ percentile measure gives a lower  estimate, therefore, we stick to the significant height $H_S$ measure to consider only the very large events as extreme events. 
The large spiking events continue with decreasing forcing frequency, however, the intermittent spikes become more and more frequent (Fig.~2, lower left panel). As a result, the nominal value becomes large as reflected by the mean value and thereby  $H_S$ (dashed horizontal line) also becomes very large such that all the spiking events remain below this threshold and we do not claim them as extreme events anymore. We stick to the definition of the $H_S$ line and ignore the  $99^{th}$ percentile measure (solid horizontal line) as an overestimation since the large events are no more rare then.    
\par  A better visual picture  of the dynamics of the Li\'enard system is presented in a series of Poincar\'e surface of section plots (Fig.~2, right column) in support of their respective time evolution (left panels) that explains the emergence of extreme events from a dormant state of bounded chaos. The Poincar\'e plot (black attractor, upper right panel) confirms chaotic dynamics with no large event that corresponds to its temporal dynamics (left panel) for a
forcing frequency $\omega=0.735$ away from the critical value.  A big closed curve (blue color) corresponds to the coexisting quasi-periodicity that limits the extent of the dissipative dynamics.
The bounded chaotic attractor (black dots)  exists in close vicinity of a saddle orbit (an open circle), otherwise, the broad state space inside the  closed curve (blue line), is empty. The trajectory of the chaotic orbit is bounded and never collides with the saddle orbit and thereby no large event appears in the temporal dynamics (upper row, left panel). The extreme events' scenario as occasional large events emerges when the forcing frequency is decreased and arrives at a  critical value $\omega=0.7315$. Extreme events are reflected in a cloud of sparsely distributed points (black dots) outside the main chaotic attractor (original black attractor of black dots) as shown in the Poincar\'e plot in Fig.~2 (middle row, right panel) that corresponds to the temporal dynamics of its left panel. The Poincar\'e plots explain the interior crisis phenomenon leading to extreme events. 
At the critical value $\omega$=0.7315, the system reaches the crisis point \cite{reinoso, celso} when the chaotic trajectory  (black dots) collides  with the saddle orbit at the origin and its stable manifold (open circle, middle right panel). As a result, the trajectory moves occasionally away from the former bounded attractor towards the boundary delineated by the closed curve (blue color) of quasiperiodic oscillations. The trajectory never  leaves the region bounded by the quasiperiodic curve (blue line), but comes back to the former bounded attractor and is repeatedly pushed away, thereby forming  recurrent and rare large amplitude spikes, which we call as extreme events. 
Such a period-doubling route to chaos followed by an  interior crisis with a sudden expansion of the attractor leading to extreme events is common in many other dynamical systems such as excitable systems \cite{karnatak} or oscillatory systems \cite{cristina, bonatto}. For smaller values, in our example system, the large size dynamics continues (Fig.~1), however, we notice very frequent large spiking events in the time evolution (lower left panel), say, for $\omega=0.7$ as an example. The mean value of $y_{max}$ becomes large and accordingly the $H_s$ line (dashed horizontal line) is even larger than all the peaks and hence no event qualifies as extreme. We consider the $99^{th}$ percentile measure (solid horizontal line) that shows as an overestimation and therefore, accept the more stringent definition of $H_s$ to include only the rarer large events. The Poincar\'e plot (lower right panel) shows the closed space within the closed curve of quasiperiodicity (blue line) filled with a galaxy of black dots corresponding to the frequent large spikes manifesting a large chaotic attractor, but no extreme event. For a further decease in $\omega$, another type of extreme events emerges which we discuss  in the next section.

\section{Extreme events: PM Intermittency} 
\begin{figure*}[]
	\hspace{-100pt}
	\begin{subfigure}[b]{0.5\columnwidth}
		\centering
		\includegraphics[width = 7.3cm,height = 4.5cm]{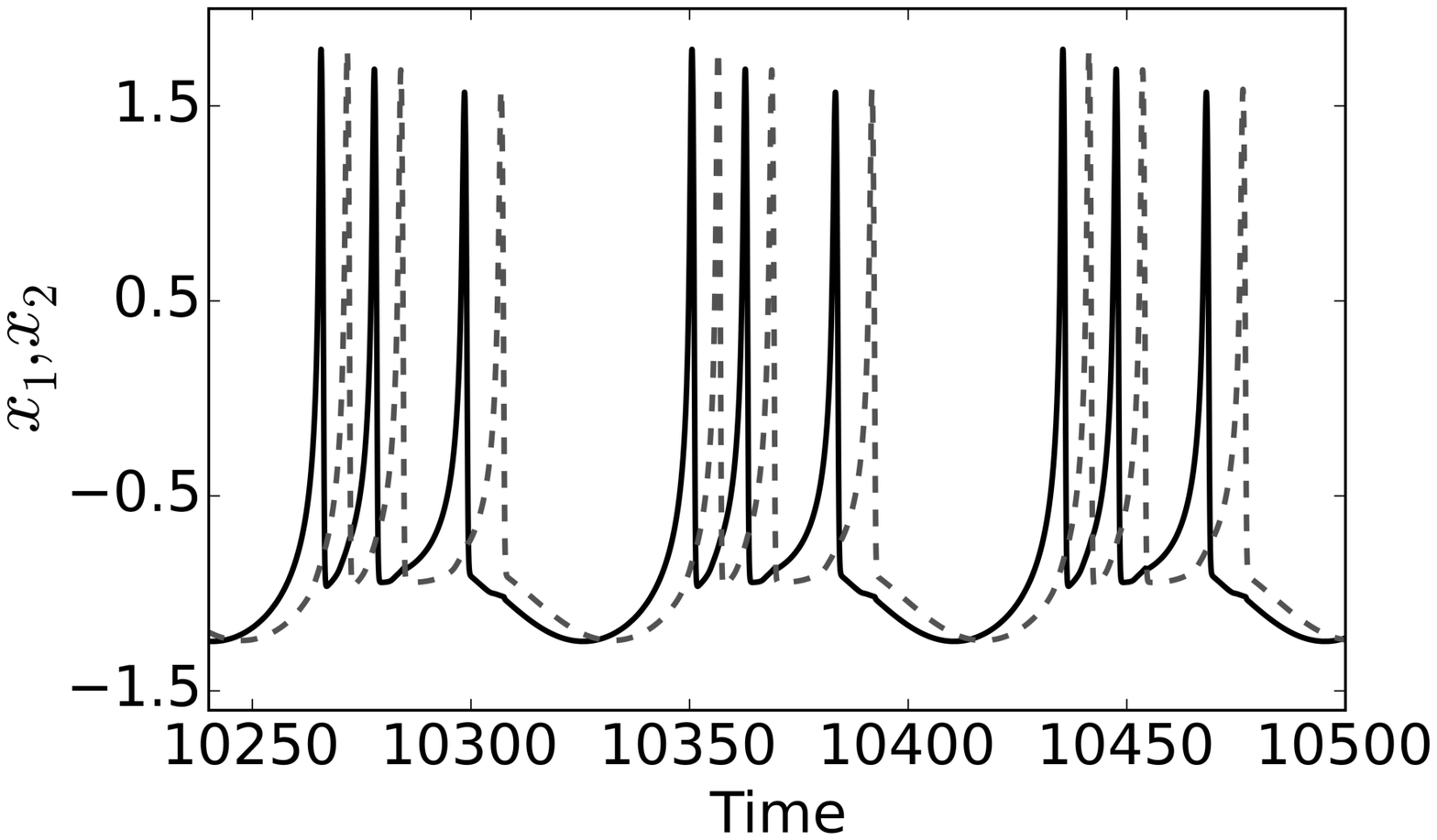}
		\label{}	
	\end{subfigure}
	\hspace{80pt}
	\begin{subfigure}[b]{0.5\columnwidth}
		\centering
		\includegraphics[width = 4.5cm,height = 4.5cm]{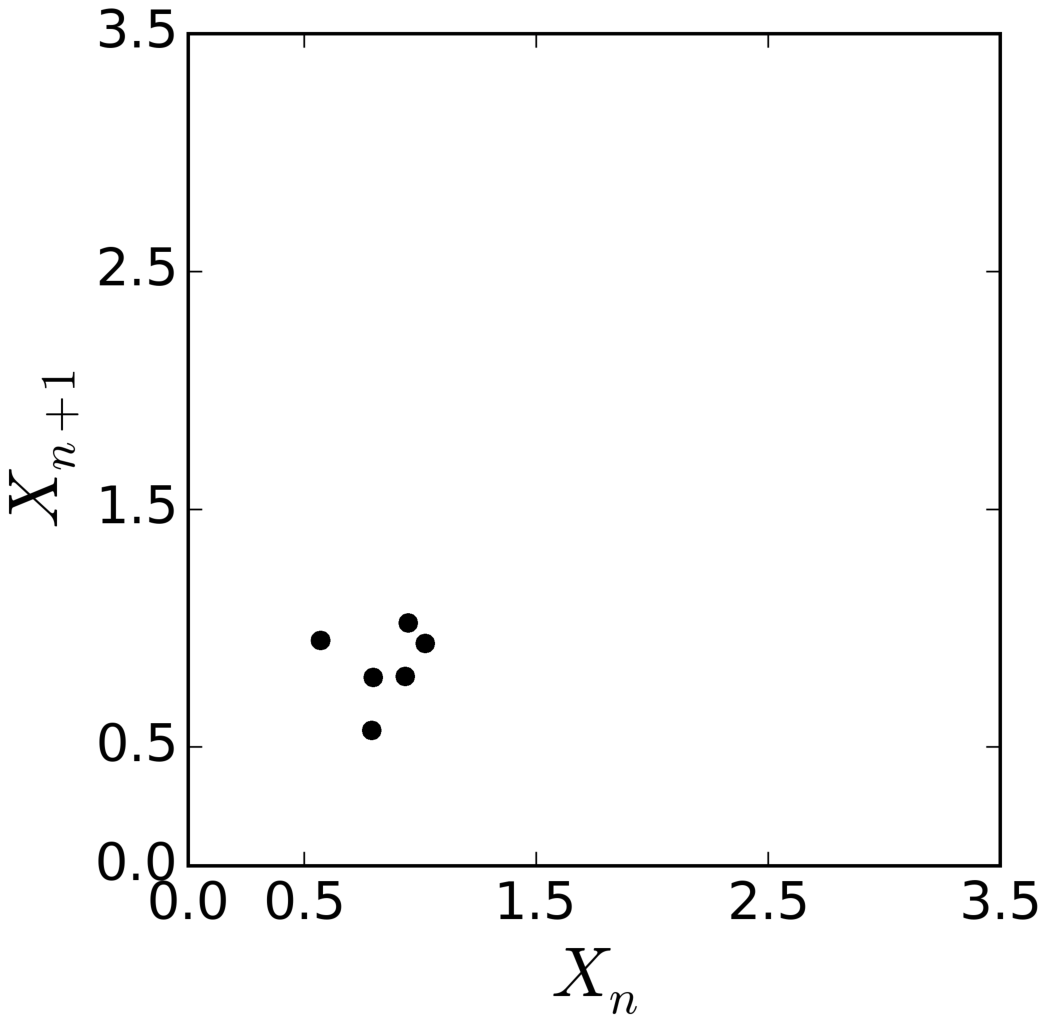}
		\label{}	
	\end{subfigure}
	\begin{subfigure}[b]{0.5\columnwidth}
		\centering
		\includegraphics[width = 6cm,height = 5cm]{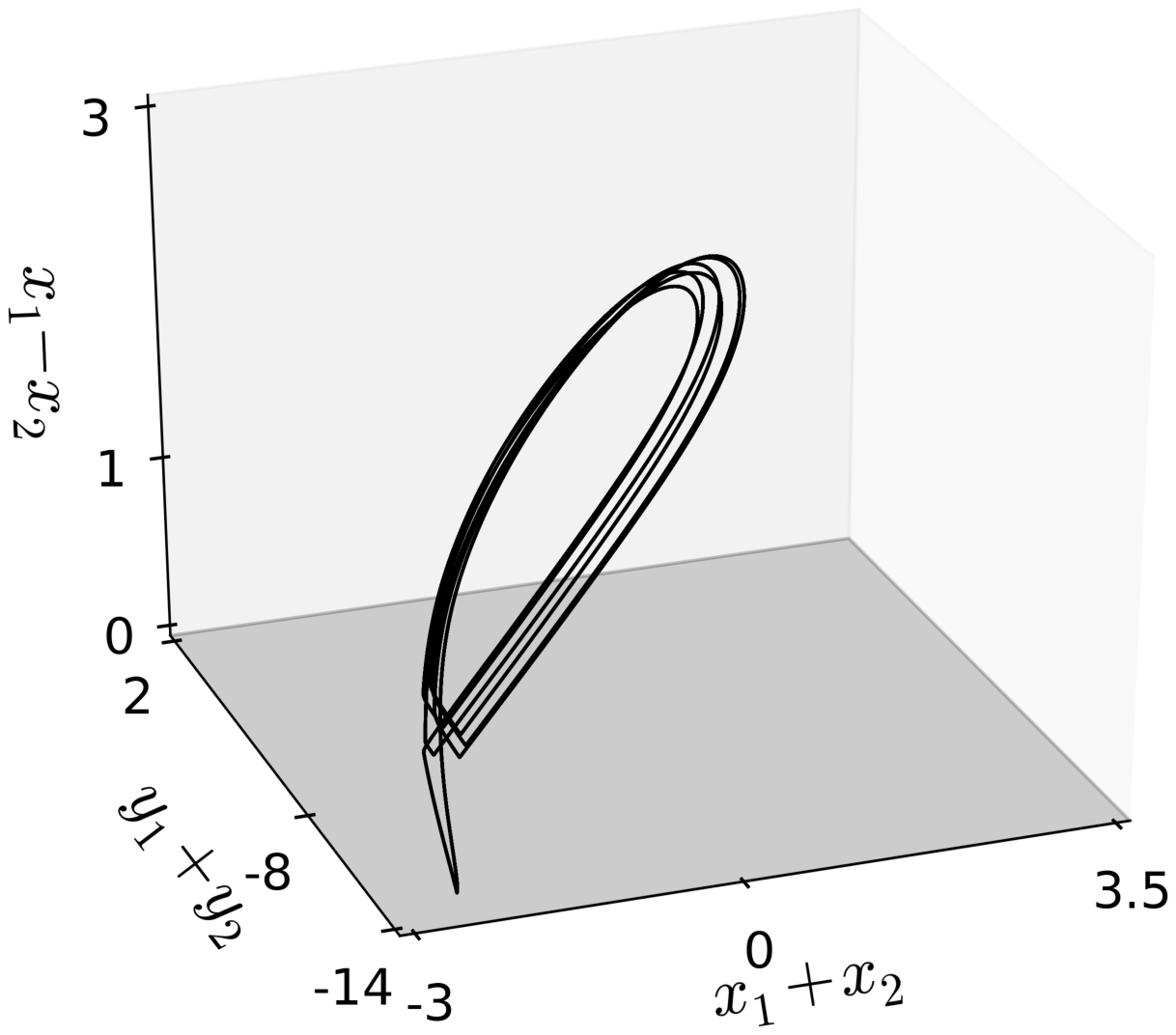}
		\label{}	
	\end{subfigure}\\
	\hspace{-100pt}
	\begin{subfigure}[b]{0.5\columnwidth}
		\centering
		\includegraphics[width = 7.5cm,height = 4.5cm]{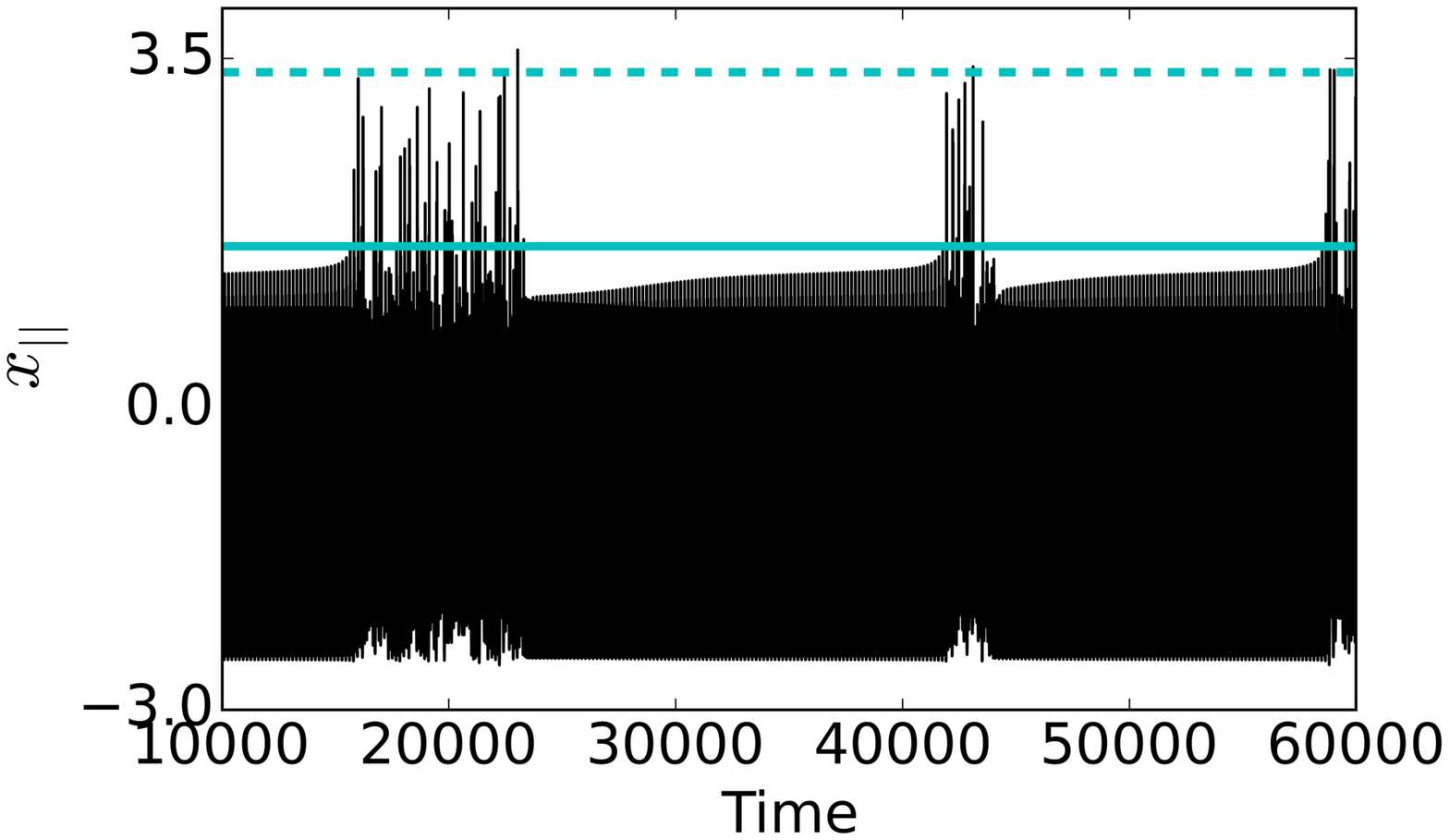}
		\label{}	
	\end{subfigure}
	\hspace{80pt}
	\begin{subfigure}[b]{0.5\columnwidth}
		\centering
		\includegraphics[width = 4.5cm,height = 4.5cm]{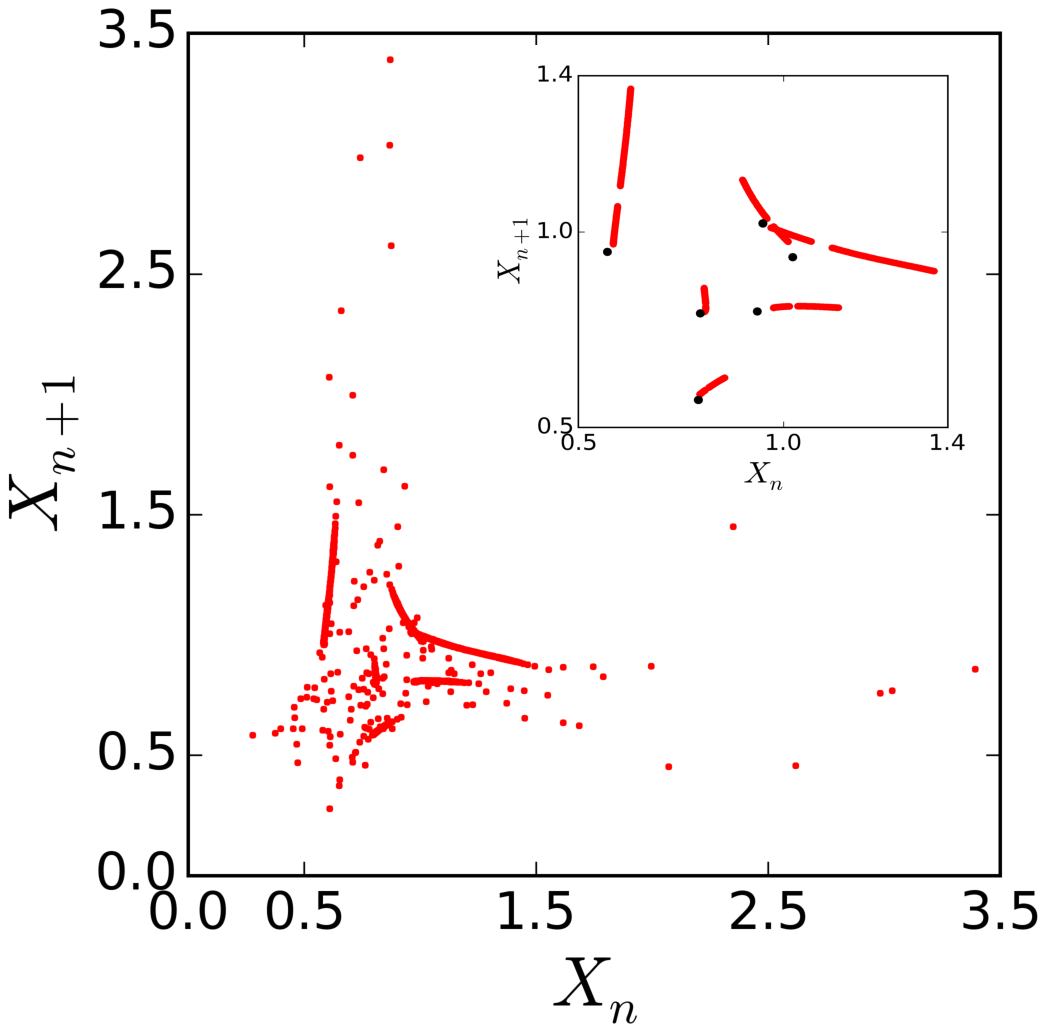}
		\label{}	
	\end{subfigure}
	\begin{subfigure}[b]{0.5\columnwidth}
		\centering
		\includegraphics[width = 6cm,height = 5cm]{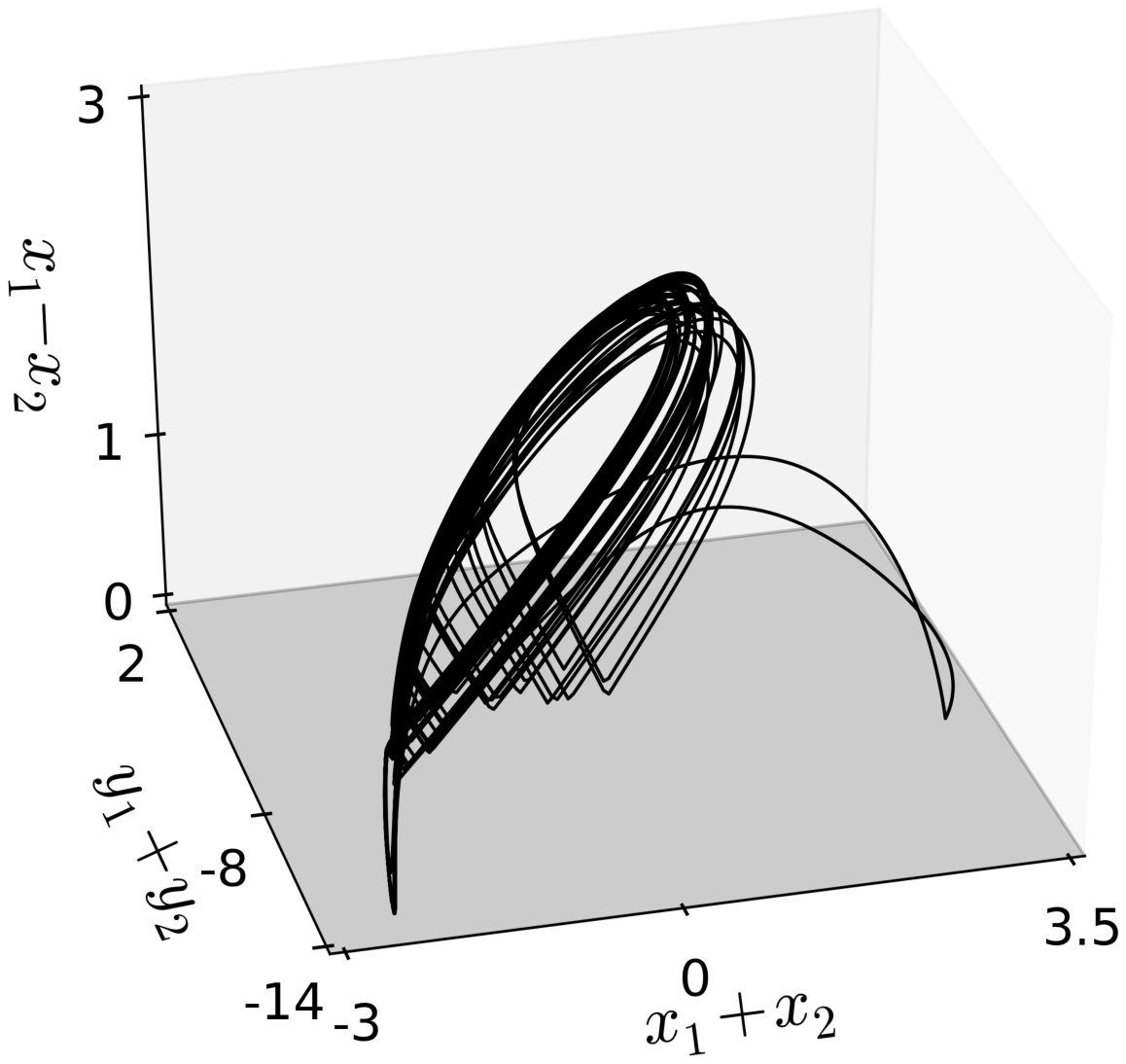}
		\label{}	
	\end{subfigure}\\
	\hspace{-100pt}
	\begin{subfigure}[b]{0.5\columnwidth}
		\centering
		\includegraphics[width = 7.5cm,height = 4.5cm]{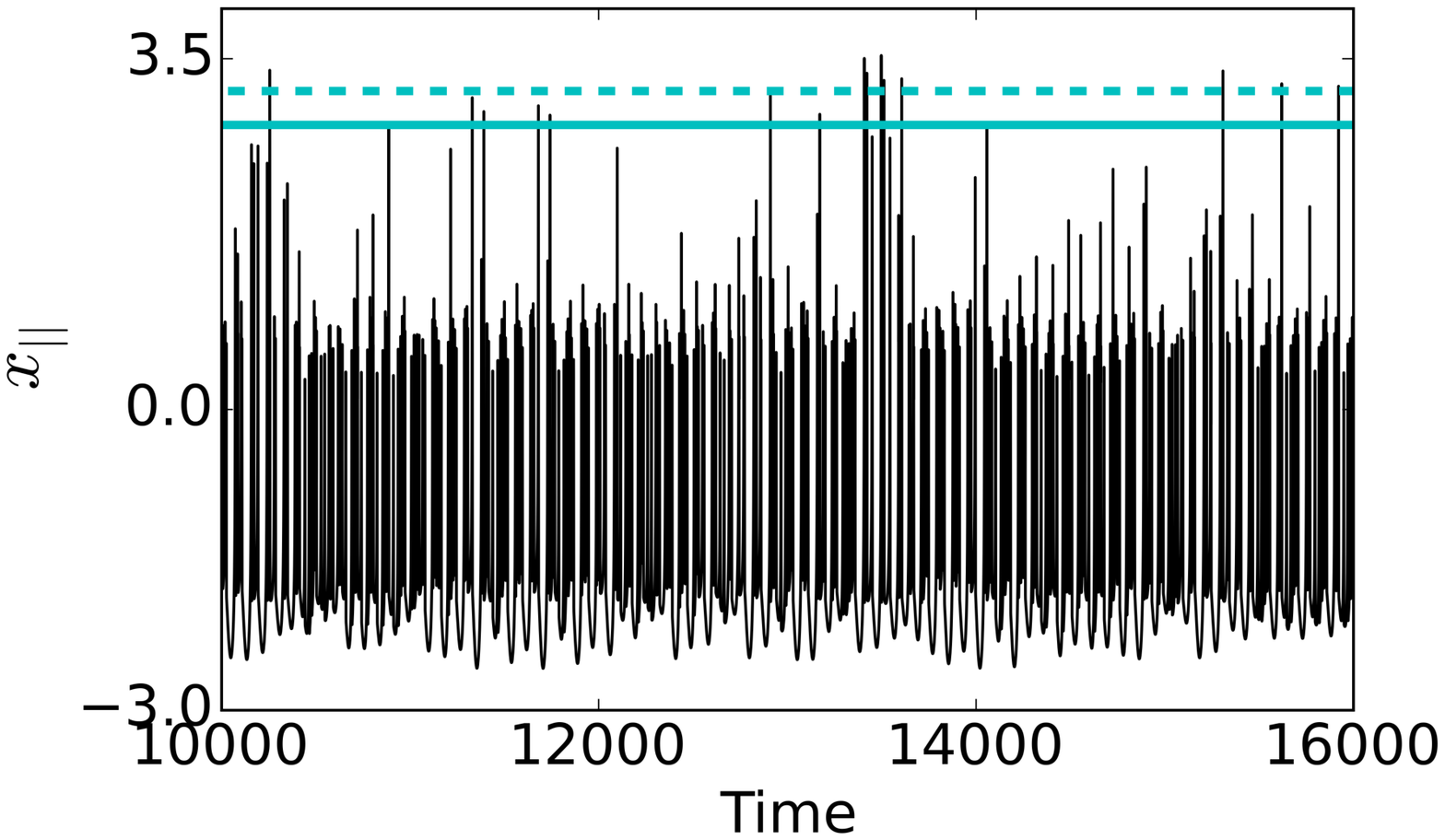}
		\label{}	
	\end{subfigure}
	\hspace{80pt}
	\begin{subfigure}[b]{0.5\columnwidth}
		\centering
		\includegraphics[width = 4.5cm,height = 4.5cm]{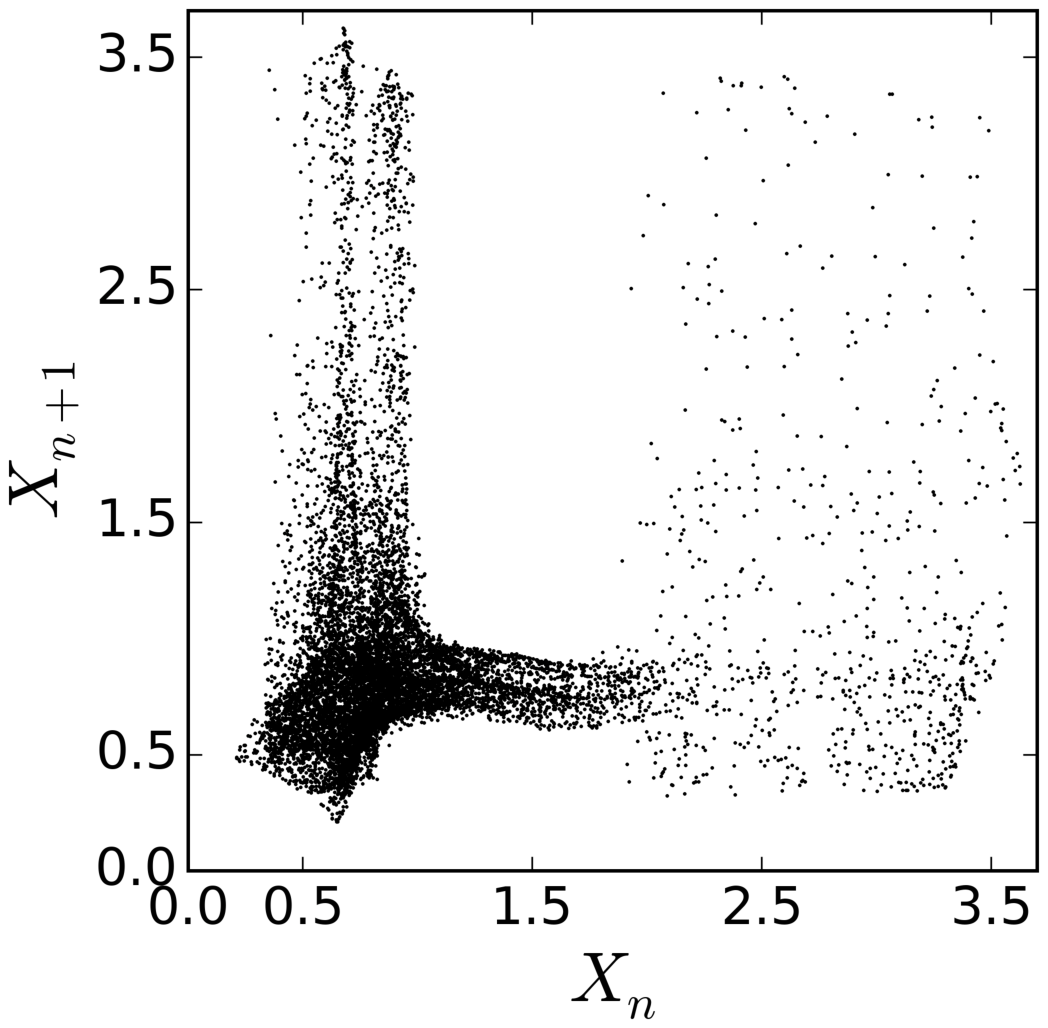}
		\label{}	
	\end{subfigure}
	\begin{subfigure}[b]{0.5\columnwidth}
		\centering
		\includegraphics[width = 6cm,height = 5cm]{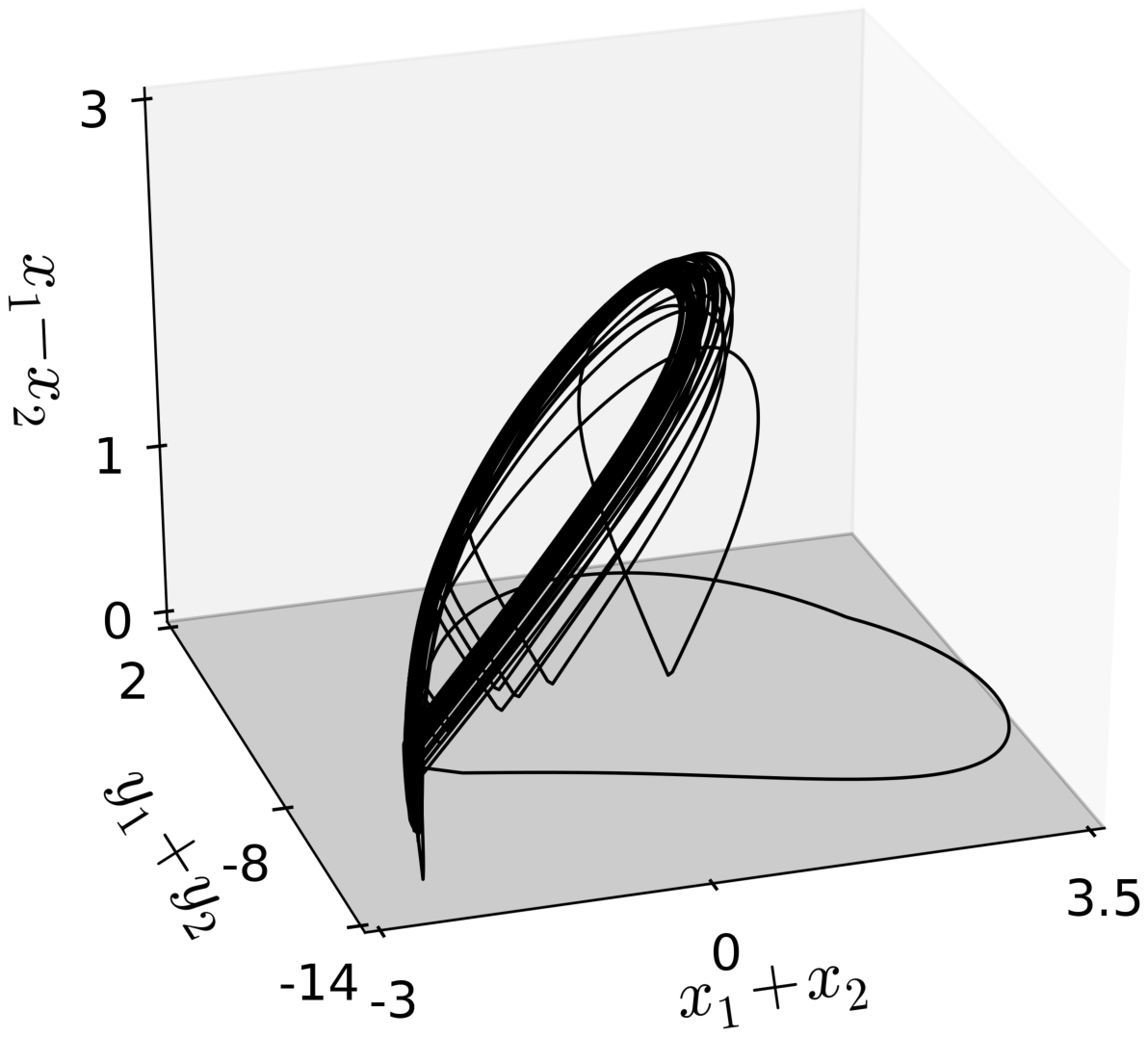}
		\label{}	
	\end{subfigure}
	\caption{(Color online) PM intermittency route to extreme events in  coupled HR model. Temporal evolution of $x_1$, $x_2$ and $x_{||}$ of two neurons under excitatory chemical synaptic interactions (left column). Periodic bursting (upper row, left panel),  intermittency (second row, left panel) and chaos (lower row, left panel) for $k_{1,2}= 0.04, 0.0532, 0.07$, respectively. Middle   and right columns correspond to their immediate left column plots. Panels in middle column correspond to the return map of $X_n$. In the second row of middle column, specifically, return map is shown colors (red). Six small color circles (blue) in the inset depict the original six fixed points as shown in the immediate upper panel (black circles). Right column denotes antispike synchronization manifold in a 3D plot of $x_{\perp}$, $x_{||}$ and  $y_{||}$.}
	\label{FIG.4}
\end{figure*}
We elaborate here the second important route to extreme events, namely, the PM intermittency. Once again, we focus on the bifurcation diagram in Fig.~1 (extreme left side) where another large increase in the amplitude of oscillations in the forced Li\'enard system occurs directly from a periodic orbit of smaller size, with increasing forcing frequency.  The temporal dynamics in Fig.~3 (upper left panel), for $\omega=0.642$, clearly shows a periodic oscillation represented by one filled circle in the Poinacar\'e plot (right panel), whereas the open circle denotes a saddle orbit.

The time evolution, at critical point value $\omega=0.6423$, is shown in Fig.~3 (lower left panel). 
It shows laminar phases of almost periodic oscillations with intermittent large spiking and  bursting, a typical characteristics of PM intermittency \cite{PM}, when a few large spikes indeed cross the horizontal $H_S$ line (dashed horizontal line) that indicates the emergence of extreme events. The $99^{th}$ percentile (solid horizontal line) measure includes all the spiking and bursting giving an overestimation of extreme events and we do not consider this measure. 
Such intermittent turbulent phases originate via PM intermittency in many systems including the Lorenz system \cite{PM}, but the heights of the spikes or bursts in a turbulent phase may not grow large enough for all systems so as to be considered as extreme. 
 The far distant scattered points in the Poincar\'e plot (Fig.~3, lower  right panel) correspond to the large events that are classified here as extreme events corroborating to their time evolution (lower left panel). 
 \par We present a second example of  PM intermittency-induced extreme events using  a coupled  model of two HR \cite{HR} neurons, 
\begin{gather*}
\dot{x_{i}} = y_{i} + bx_{i}^{2} - ax_{i}^{3} - z_{i} + I - k_{i}(x_{i} - v_{s})\Gamma(x_{j})\\
\dot{y_{i}} = c - dx_{i}^{2} - y_{i}\\
\dot{z_{i}} = r[s(x_{i} - x_{R}) - z_{i}],
\end{gather*}
where $i, j=1, 2$($i\neq j$) denote two oscillators, $\Gamma(x)$=$\frac{1}{1 + \exp^{-\lambda(x-\Theta)}}$ is a sigmoidal function 
that represents the chemical synaptic interaction between the neurons\cite{HR}.
The parameters of the coupling function are  $v_{s}$ = 2, $\lambda$ = 10, $\Theta = -0.25$.  The coupling constant $k_{1,2}$  is taken  positive to keep the mutual communication between the neurons always excitatory.
We start with two periodically bursting neurons (for the dynamics of a single neuron including details about the parameters see Appendix A), as a case study, which remain locked either in antispike or antiburst synchronization for small coupling (attractive or repulsive synaptic interaction, see \cite{mishra} for details).
\par For a complete description of our results, we draw three sets of plots in Fig.~4, the time evolution of the coupled  dynamics (left column), return maps (middle column) of $X_n=max(x_{||})$ where $x_{||}=(x_1+x_2$) and a projection of the antiphase synchronization manifold in a 3-dimensional (3D) space (right column) of $x_{\perp}$=($x_1-x_2$) defining the transverse direction to the in-phase manifold and a plane (dark grey) of in-phase manifold, $x_{||}(=x_1+x_2$) vs. $y_{||}=(y_1+y_2$). For a weak attractive coupling ($k_1=k_2=k=0.04$), the dynamics of the coupled system  remains  periodic (bursting) and in stable antispike synchronization as shown in the temporal dynamics of the two oscillators $x_1$ (solid lines) and $x_2$ (dashed line) in Fig.~4 (upper left panel) where the spikes of two bursts show a lag. We capture all the maxima of the time evolution of $max(x_{||})=X_n$ and plot them in a $X_n$ vs. $X_{n-1}$ return map (middle column), each corresponding to its immediate left panel. The phase locking of the coupled oscillators  is revealed in the return map (six closed circles) that corresponds to six collective spikes (upper left panel), three spikes for each oscillator lagging behind each other. The 3D plot (right column) confirms a stable phase locking.
Instabilities in antispike synchronization arise with increasing coupling strength as shown in Fig.~4 (middle row) via PM intermittency for $k=0.0532$. 
Intermittent bursting events are seen in the time evolution of $x_{||}$ (middle row, left panel) between the  laminar phases (only a few bursts are shown, however, in a long time series, it recurs many times). The bursts in the turbulent phases show high amplitude spikes, a few of them cross the  $H_S (=\mu+6\sigma$) line (dashed horizontal line) indicating rare occurrence of extreme events. The $99^{th}$ percentile line (solid horizontal line) again overestimates the extreme events. 
The origin of complexity with coupling is more prominent in the return maps of $X_n$. 
The intermittency emerges with an instability in the time evolution of $x_{||}$ (second row, left panel), which is reflected in the return map as elongated lines (second row, middle panel, six broken red lines) along with scattered points (red dots) indicating large amplitude spikes. The elongated lines correspond to the laminar phase with slowly increasing amplitude as elaborated in the inset where a map is drawn  using only the data from the laminar phase. The phase locking is  maintained in the laminar phase, but breaks intermittently during the turbulent phase.  On the other hand, the coexisting scattered points (red dots) are seen at far away locations, corresponding to the large events.  The 3D trajectory exhibits, most of the time, an antiphase synchronization (second row, right panel), but also shows a rare journey towards the in-phase manifold due to a local  instability. At a larger coupling ($k=0.07$), the temporal dynamics of $x_{||}$ becomes more complex with frequent spiking. Both the $H_S$ and $99^{th}$ percentile measures almost converge, in this case, when few spikes rarely cross the horizontal lines, indicating the origin of extreme events. The return map (third row, middle panel) becomes dense along with a cloud of scattered points on the right without showing any stable structure. The antispike  synchronization is unstable (lower right panel) and an increasing number of trajectories moving along the in-phase synchronization manifold (one exemplary 3D trajectory is shown, for a short time duration). For further increase in coupling, very frequent spikes appear (not shown here), which do not qualify as extreme events for reasons discussed in the previous example. Note that PM intermittency route has not been discussed in the context of  extreme events, in  literature, except that its possibility has been mentioned earlier \cite{nicolis} from a statistical point of view and very recently, in an experiment of a thermoacoustic system \cite{sujith}. 
\section{Extreme events: Quasiperiodic route}
\begin{figure*}[]
	\hspace{-100pt}
	\begin{subfigure}[b]{0.5\columnwidth}
		\centering
		\includegraphics[width = 7cm,height = 4cm]{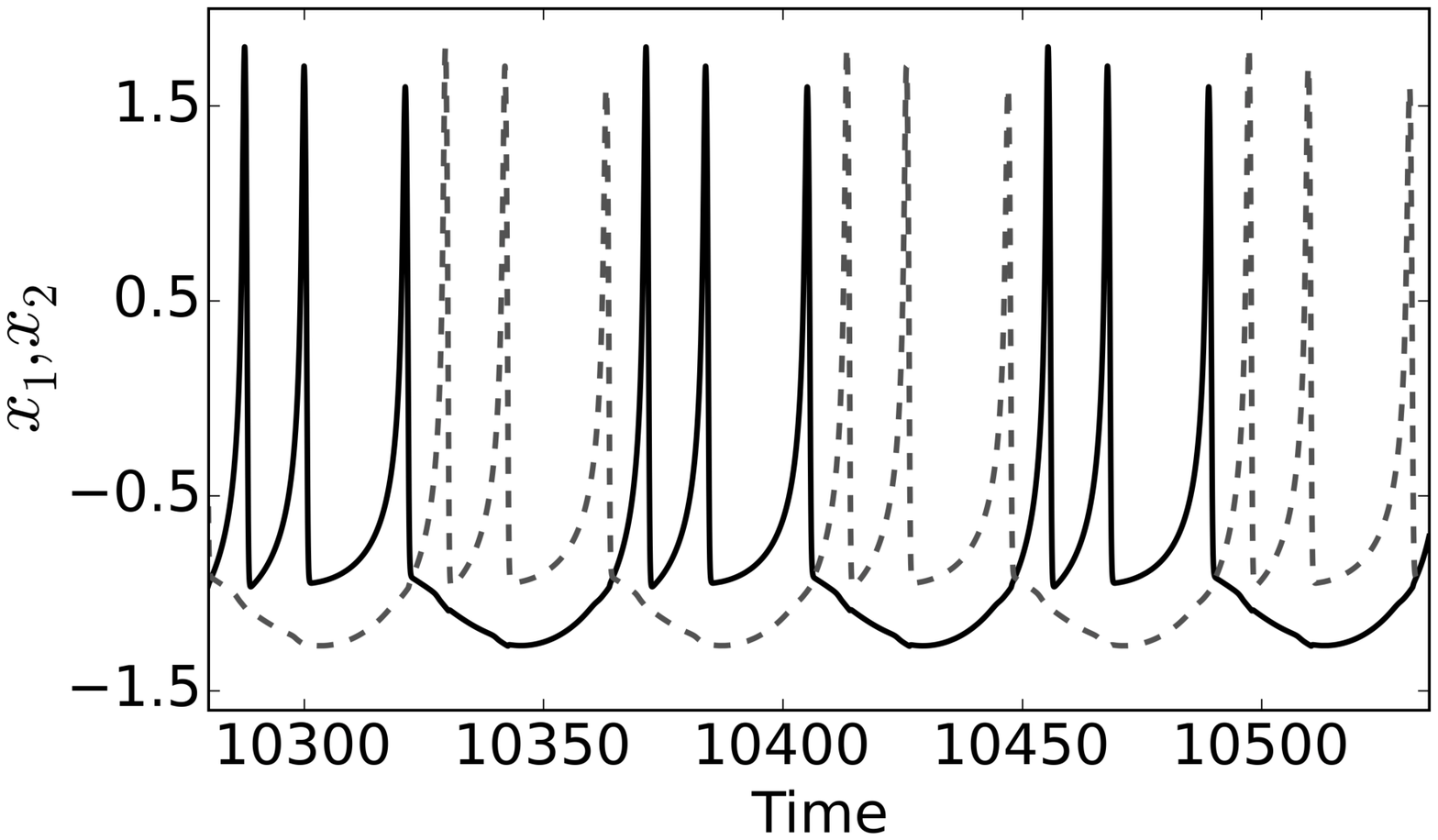}
		\label{}	
	\end{subfigure}
	\hspace{70pt}
	\begin{subfigure}[b]{0.5\columnwidth}
		\centering
		\includegraphics[width = 4.75cm,height = 4.15cm]{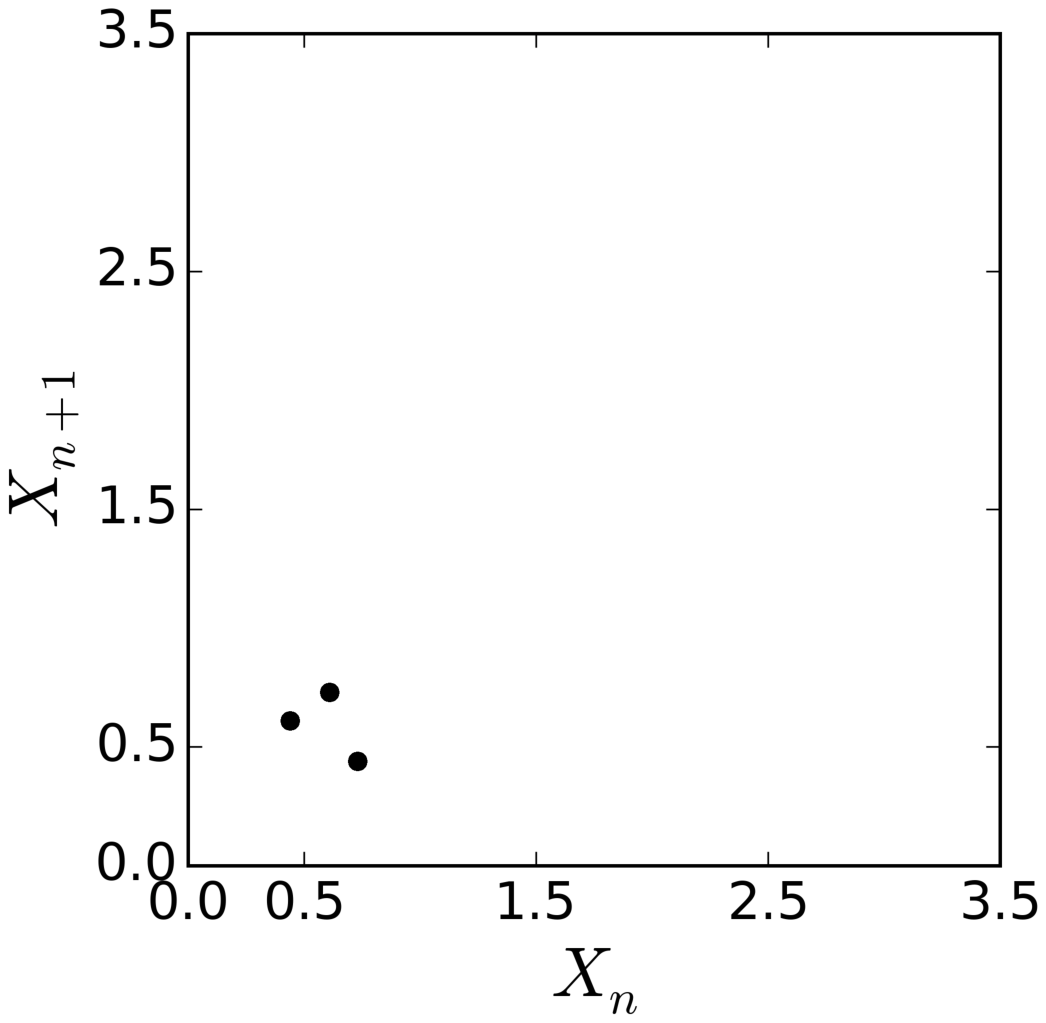}
		\label{}	
	\end{subfigure}
	\hspace{0pt}
	\begin{subfigure}[b]{0.5\columnwidth}
		\centering
		\includegraphics[width = 6cm,height = 5cm]{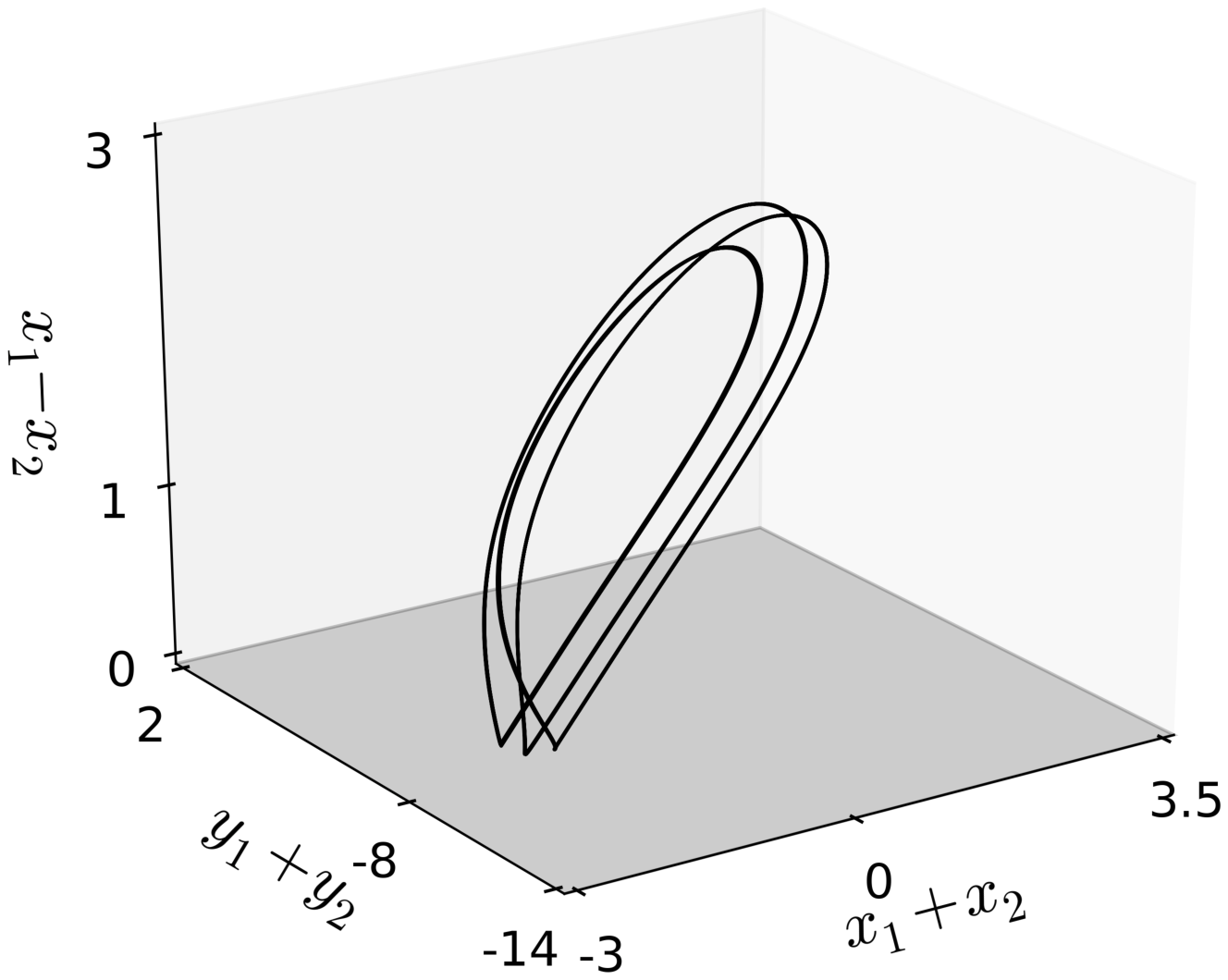}
		\label{}	
	\end{subfigure}\\
	\vspace{-20pt}
	\hspace{-100pt}
	\begin{subfigure}[b]{0.5\columnwidth}
		\centering
		\includegraphics[width = 7.25cm,height = 4cm]{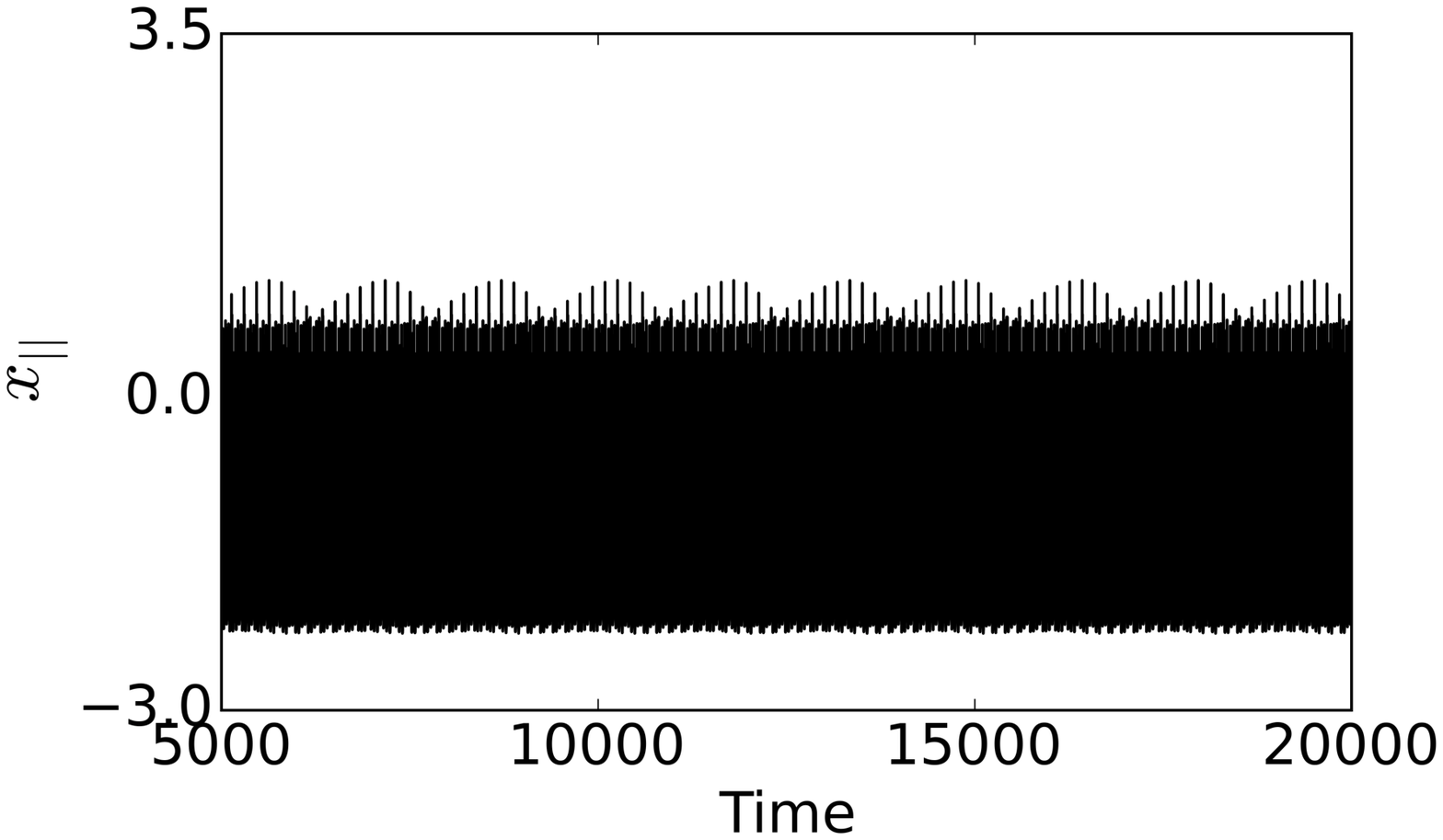}
		\label{}	
	\end{subfigure}
	\hspace{70pt}
	\begin{subfigure}[b]{0.5\columnwidth}
		\centering
		\includegraphics[width =4.75cm,height = 4cm]{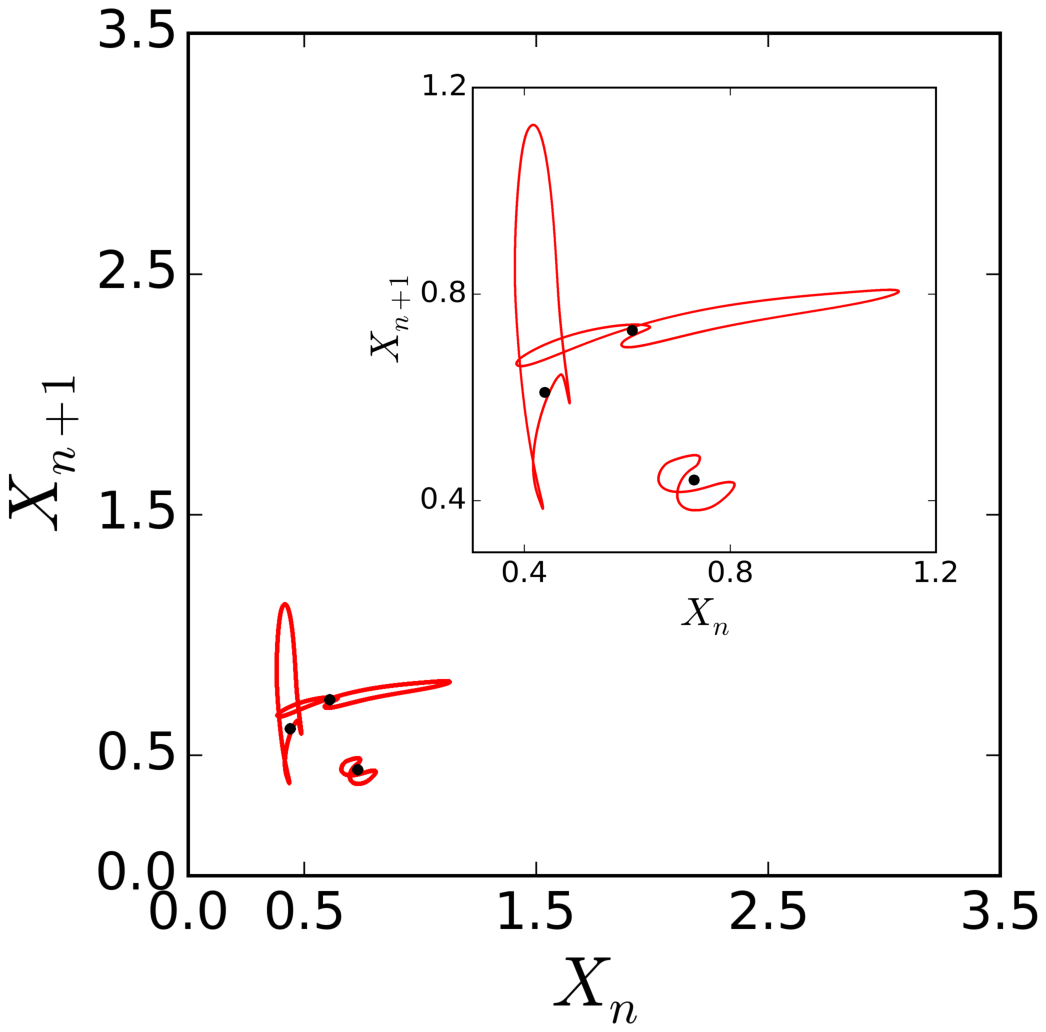}
		\label{}	
	\end{subfigure}
	\hspace{0pt}
	\begin{subfigure}[b]{0.5\columnwidth}
		\centering
		\includegraphics[width = 6.25cm,height = 5cm]{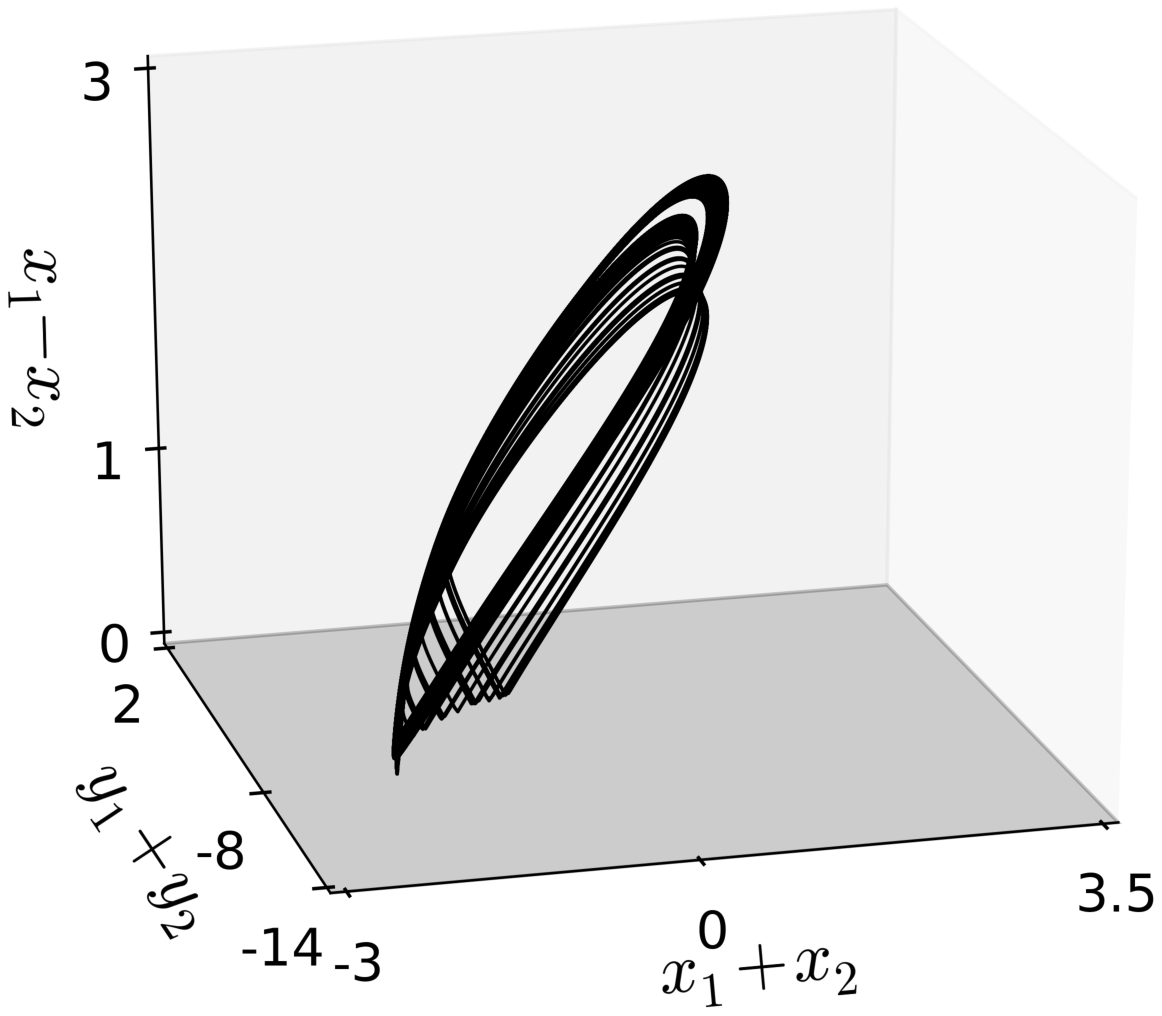}
		\label{}	
	\end{subfigure}\\
	\hspace{-100pt}
	\begin{subfigure}[b]{0.5\columnwidth}
		\centering
		\includegraphics[width = 7.25cm,height = 4cm]{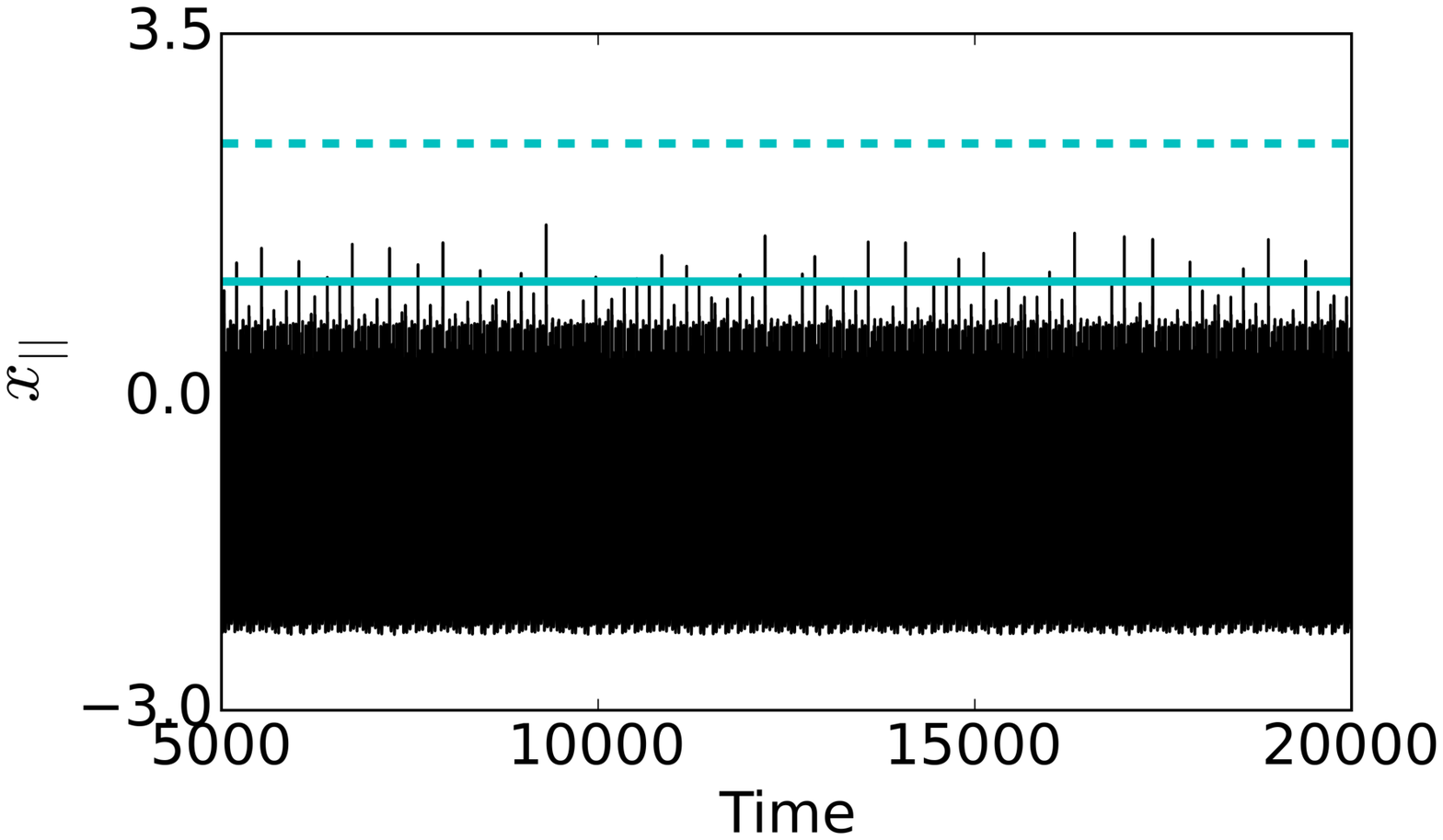}
		\label{}	
	\end{subfigure}
	\hspace{70pt}
	\begin{subfigure}[b]{0.5\columnwidth}
		\centering
		\includegraphics[width = 4.75cm,height = 4cm]{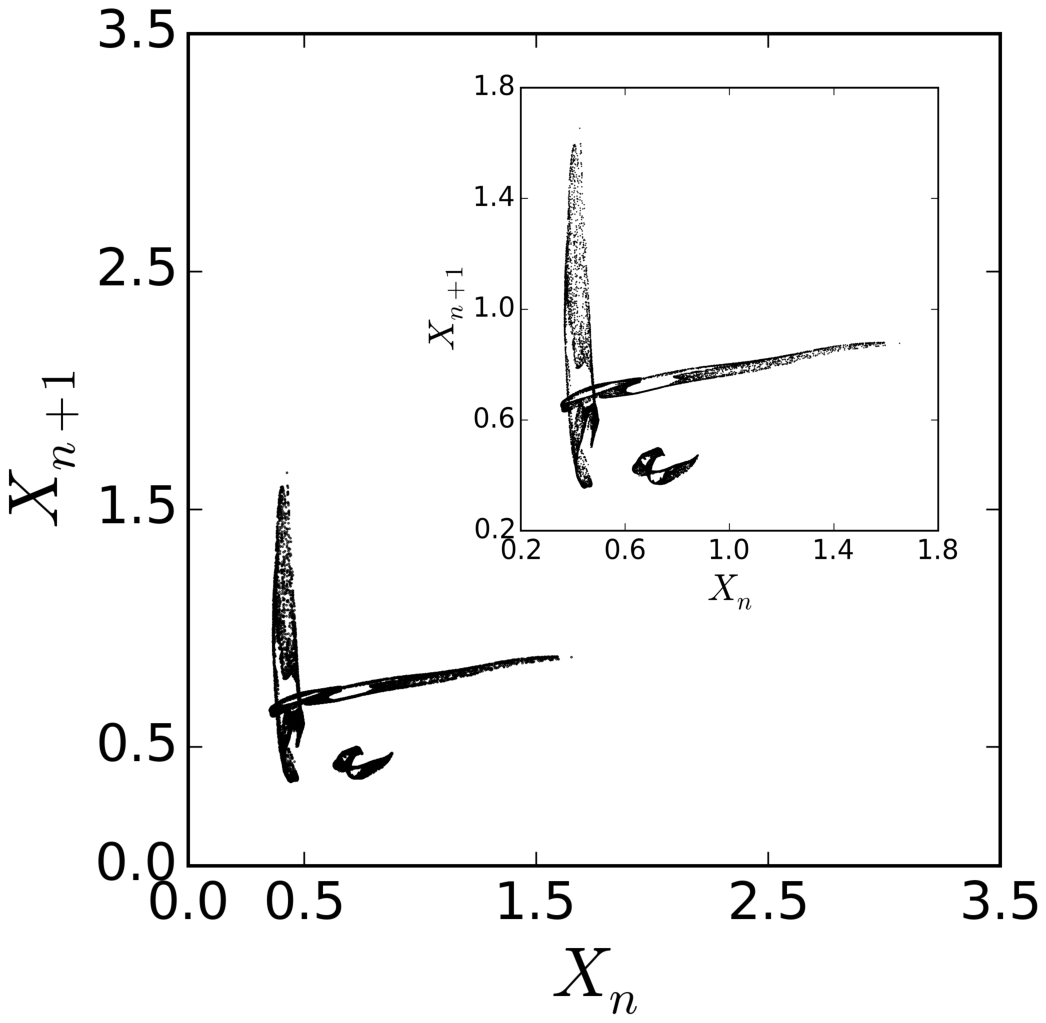}
		\label{}	
	\end{subfigure}
	\hspace{10pt}
	\begin{subfigure}[b]{0.5\columnwidth}
		\centering
		\includegraphics[width = 6cm,height = 4.5cm]{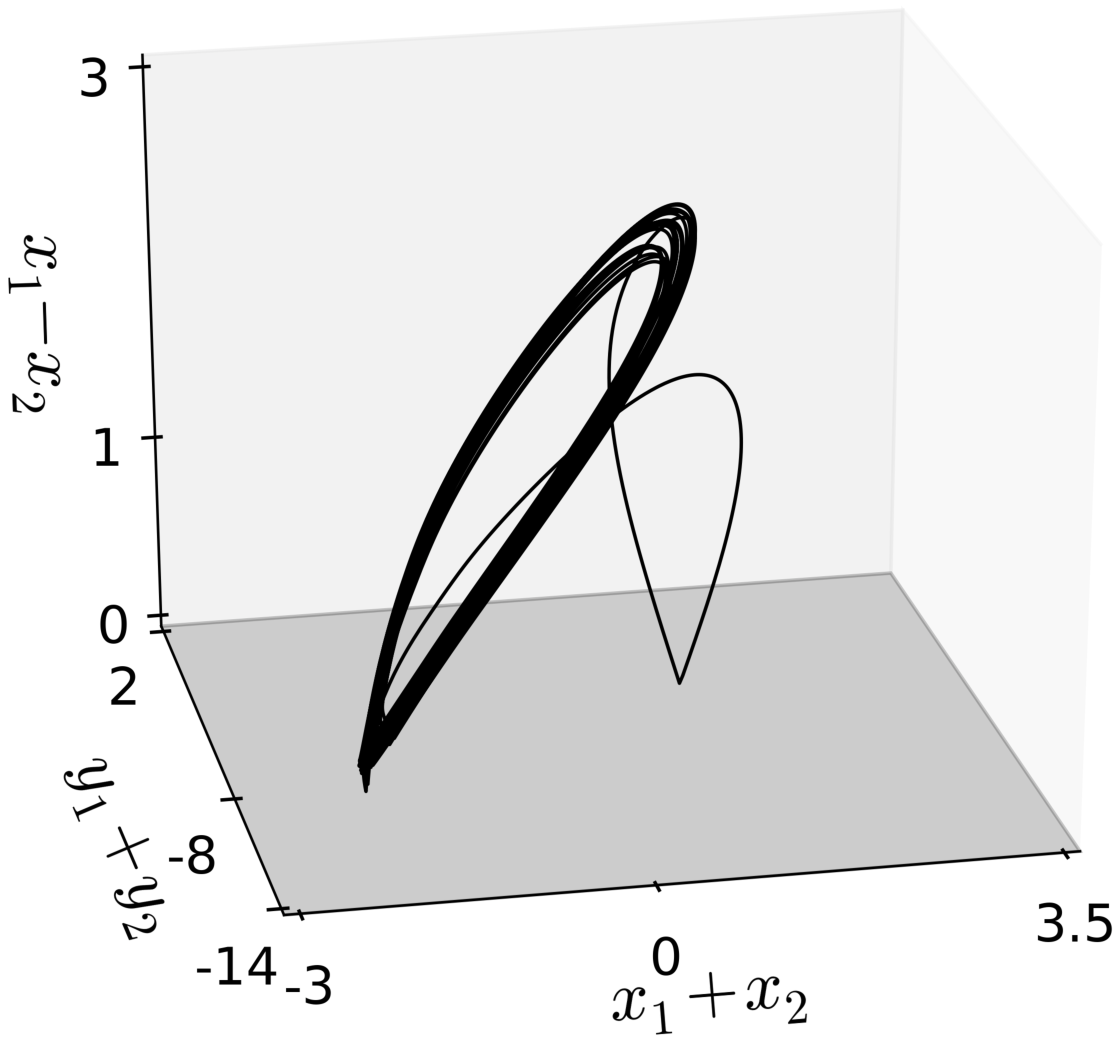}
		\label{}	
	\end{subfigure}\\
	\hspace{-90pt}
	\begin{subfigure}[b]{0.5\columnwidth}
		\centering
		\includegraphics[width = 6.75cm,height = 4cm]{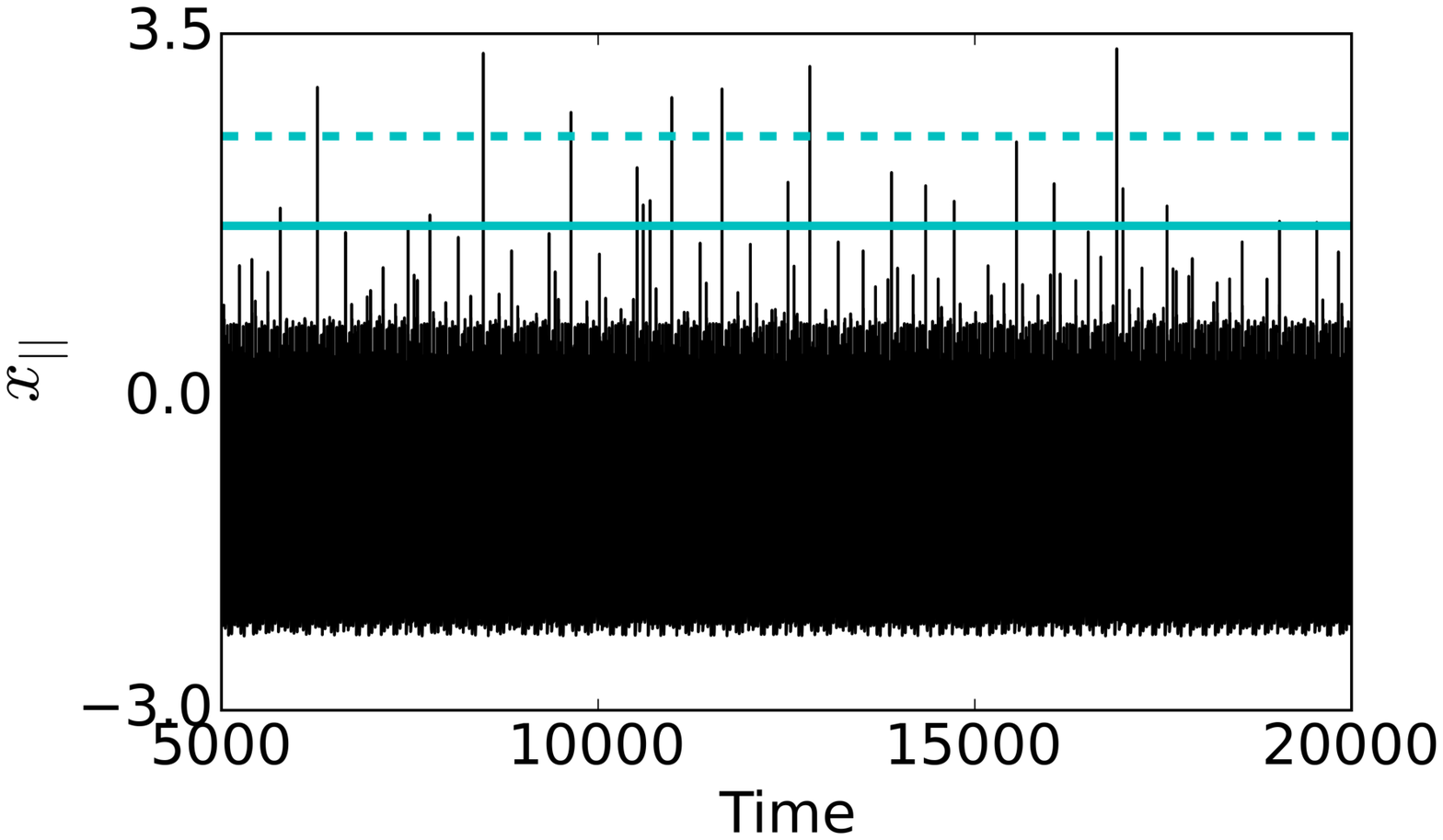}
		\label{}	
	\end{subfigure}
	\hspace{60pt}
	\begin{subfigure}[b]{0.5\columnwidth}
		\centering
		\includegraphics[width = 4.75cm,height = 4cm]{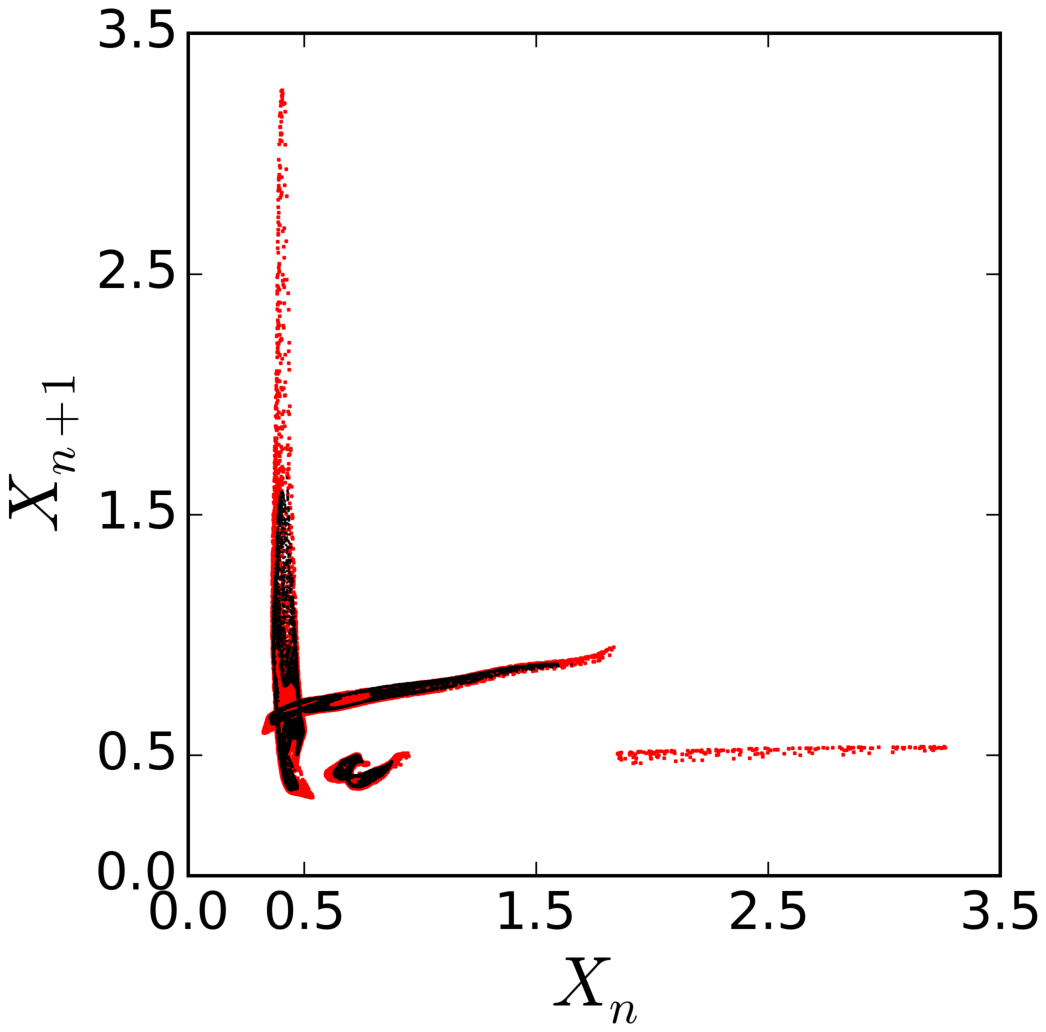}
		\label{}	
	\end{subfigure}
	\hspace{10pt}
	\begin{subfigure}[b]{0.5\columnwidth}
		\centering
		\includegraphics[width = 6cm,height = 4.5cm]{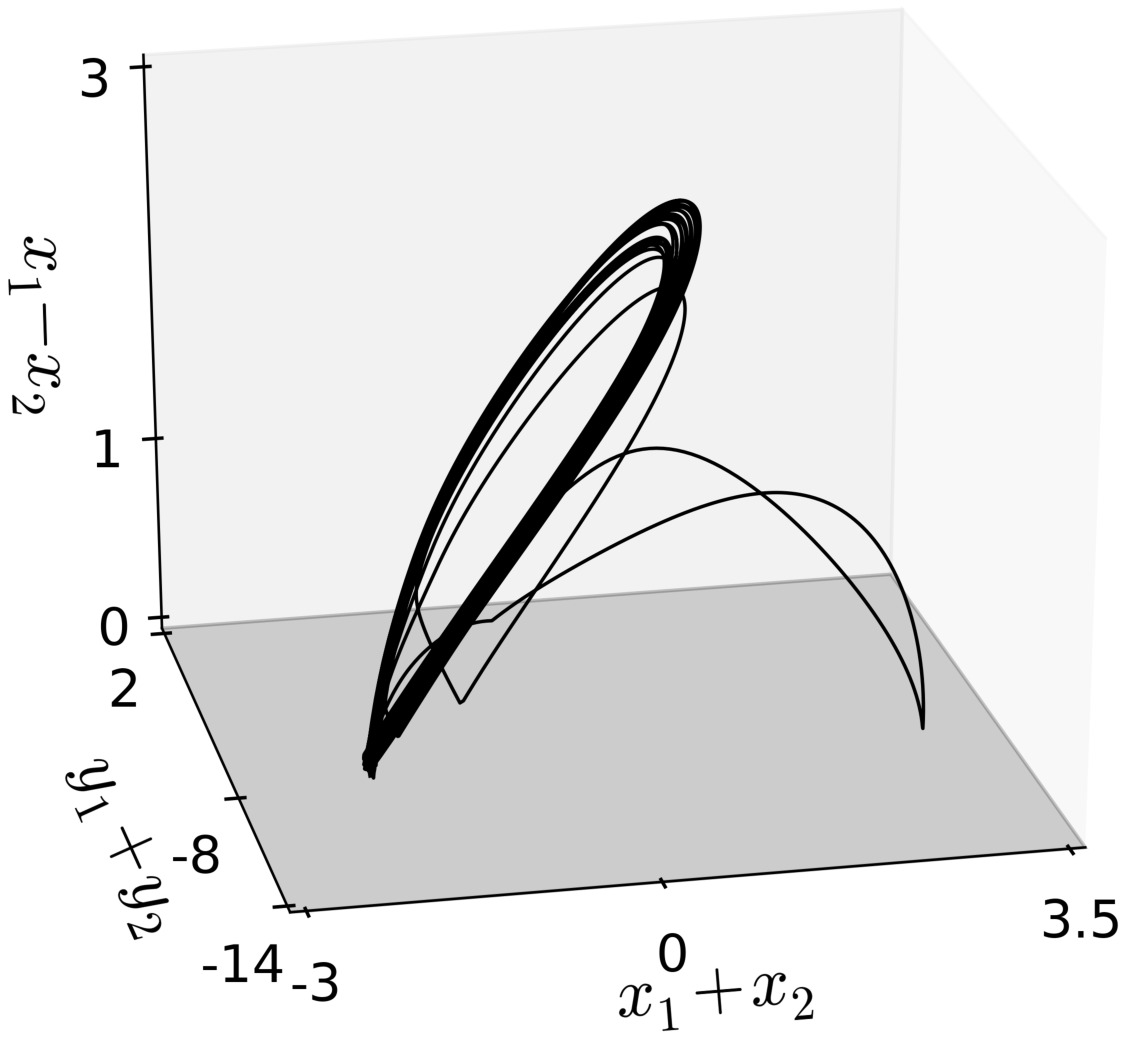}
		\label{}	
	\end{subfigure}
	\caption{(Color online) Quasiperiodic route to extreme events in the coupled HR neuron model. Temporal dynamics of $x_1$, $x_2$ and $x_{||}$  with increasing chemical synaptic interactions (left column). Sequence of events for periodic bursting (upper row),  quasiperiodicity (second row), breakdown of quasiperiodicity to chaos (third row) and extreme events (last row) for $k_{1,2}= -0.04, -0.06,-0.0686, -0.07$, respectively. The panels in the middle column  are return maps of $X_n$ corresponding to their time evolution at the left panels. The second row middle column shows the return maps of quasiperiodicity as three closed curves (red) that evolves from  three small circles (blue) drawn to depict three spikes in a periodic burst as shown in the immediate upper panel. An enlarged view in an inset shows the three small circles (black) from where the three curves (red) evolve. The lower middle panel simply shows the return map in color (red) on top of which a return map (black) of the normal chaotic case (shown in immediate upper panel) is drawn. Panels in the right column correspond to the antiburst synchronization scenario in a 3D plot of $x_{\perp}$, $x_{||}$ and  $y_{||}$. }
	\label{FIG.5}
\end{figure*}
A third kind of scenario, namely, a quasiperiodic route to extreme events is illustrated here. 
The possibility of this route was also indicated earlier \cite{nicolis} in the case of a map, but never elaborated from a dynamical point of view.  To investigate this phenomenon, we employ the same example of the coupled neuron model. This time the synaptic interaction between the two neurons is considered as inhibitory instead of excitatory \cite{mishra}, which is realized by taking $k_{1}=k_{2}=k$ negative. We vary the mutual inhibitory interaction and once again  plot, time evolutions (left column), return maps (middle column) and a specific 3D representation of the synchronization manifold (right column) as done for the previous case. For a weak inhibition ($k=-0.04$), periodic bursting is seen as well in the time evolution plots of $x_1$ and $x_2$ in Fig.~5  (upper left panel), but it shows antiburst synchronization when each burst of three spikes of each coupled oscillator appears alternately. Three fixed points (black circles) are seen for three spikes in each burst in the return map of $X_n$ vs. $X_{n-1}$ plot (upper row, middle panel). A corresponding 3D plot of stable antiburst synchronization is shown (upper row, right panel). For stronger inhibition ($k=-0.06$), the time evolution of $x_{||}$ transits to quasiperiodicity (second row, left panel). The return map of $X_n$ vs. $X_{n-1}$ (second row, middle panel) demonstrates quasiperiodicity in the dynamics by showing three closed curves (red lines), which originated from the three fixed points (three small black circles for three spikes in a periodic burst) that are placed inside the three closed curves for a comparison (the inset presents a zoomed version). The 3D plot (second row, right panel) does not change much compared to the anti-burst state of synchrony.   For a further increase in the inhibitory synaptic interaction, a transition to chaos occurs in the dynamics via a breakdown of quasiperiodicity (third row, left panel). The $H_S (=\mu+6\sigma$) line (dashed horizontal line) is far above all the peaks and we ignore again the $99^{th}$ percentile measure for reasons discussed above. We do not consider the spikes as extreme, which is clear from the return map (third row, middle panel) where no significant expansion of the attractor is seen.  However, the closed curves of quasiperiodicity are all filled (black points) indicating the origin of chaos. The chaotic phase shows  origin of instability in the 3D plot (third row, right panel) with an occasional excursion from the antiburst synchronization manifold. The departures from the antiburst synchronization manifold are  not large enough to call them  as extremes. 

Finally, for a stronger inhibition, $k=-0.07$, extreme events start appearing as reflected by large spiking events (bottom row, left panel) some of them cross  $H_S$ line (horizontal dashed line). We consider only the largest events and ignore the $99^{th}$ percentile measure. The return map of $x_{||_{max}}$ (bottom row, middle panel) reveals a complete disappearance of three closed curves which are now fractal sets (red dots) with large enhancement of the  size. For a comparison of the size of the attractor after the expansion, the return map of the chaotic bounded state (black dots) is superimposed here. The antiburst synchronization (lower row, right column) is unstable to originate large excursions from the antiphase manifold such that it is occasionally intercepted by in-phase events. This explains our observation of the emergence of extreme events via the breakdown of quasiperiodicity.

\section{Probability distribution of events}
Finally, we discuss the statistical properties of the dynamics during  three scenarios leading to the emergence of extreme events. The probability distribution of extreme events and their return time has been estimated. The return time of extreme events is estimated by calculating the inter-event intervals that  measure the time intervals between two successive  events that qualify as extreme when crossing  $H_S$ line.  For  Li\'enard system, we draw the probability distribution function (PDF) against all events of heights ($y_n$) before and after the transition to extreme events (Fig.~6).

\begin{figure}[hbt!]
	\includegraphics[width=7.5cm, height=5.5cm]{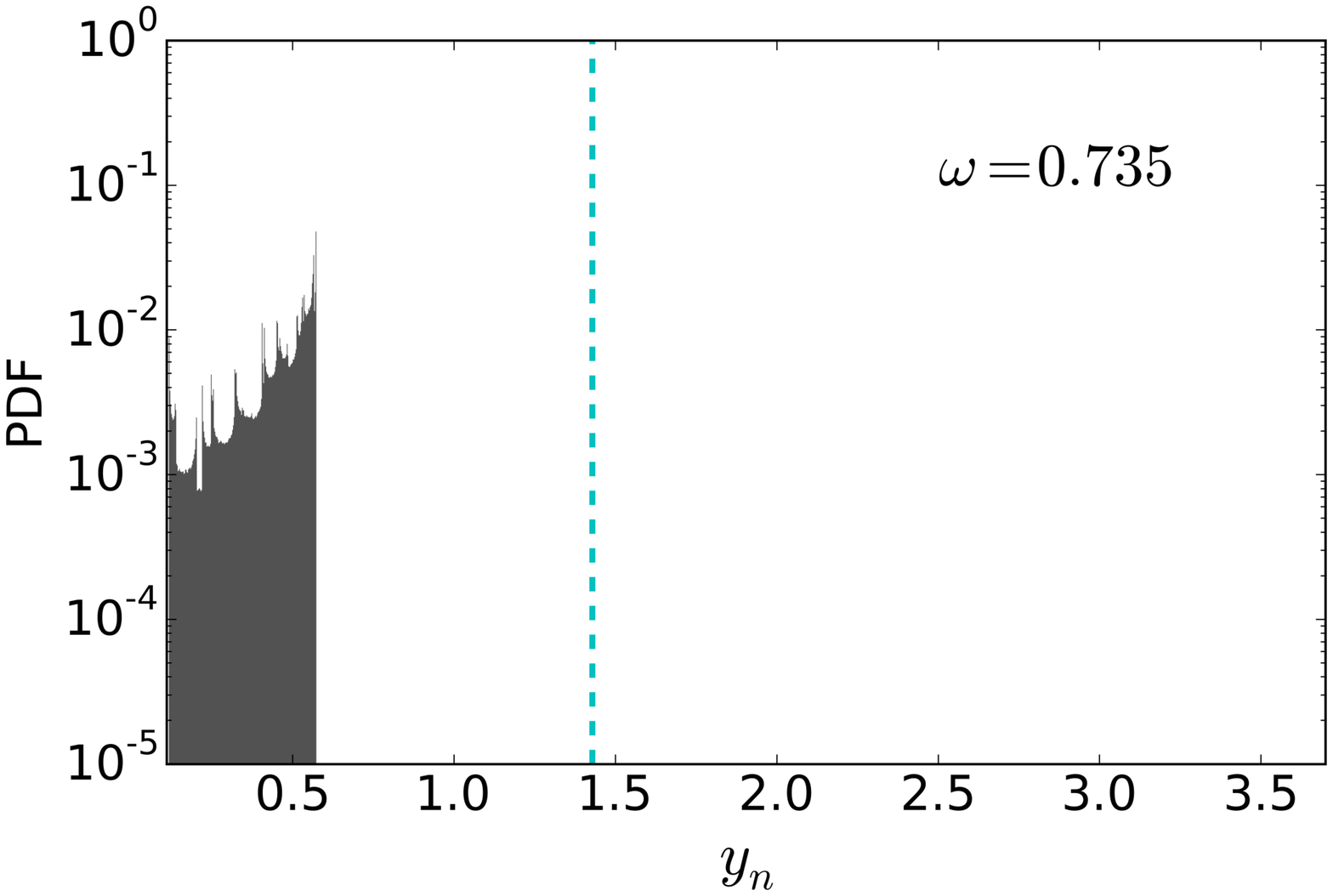}\\
	\includegraphics[width=7.5cm, height=5.5cm]{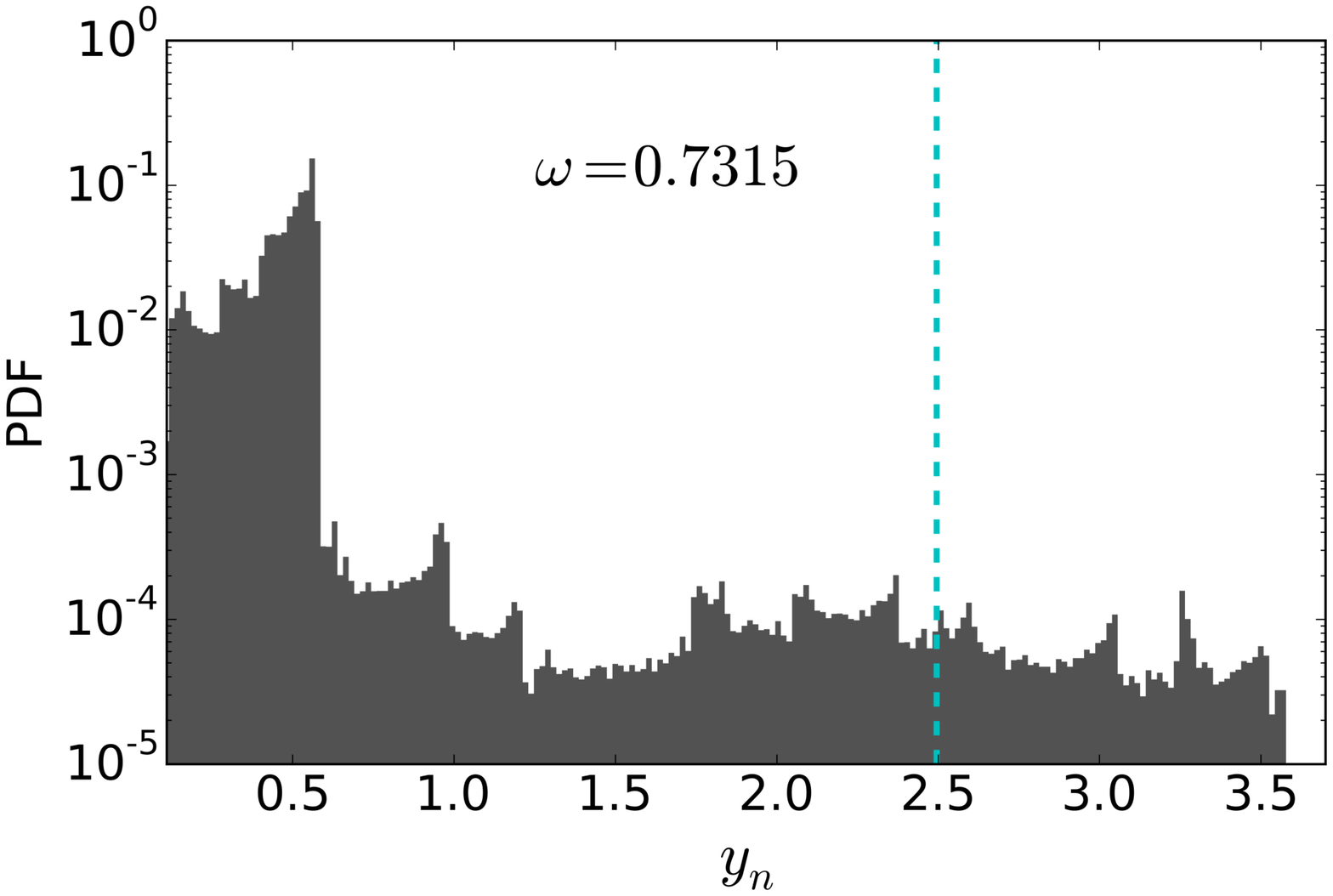}\\
	\includegraphics[width=7.5cm, height=5.5cm]{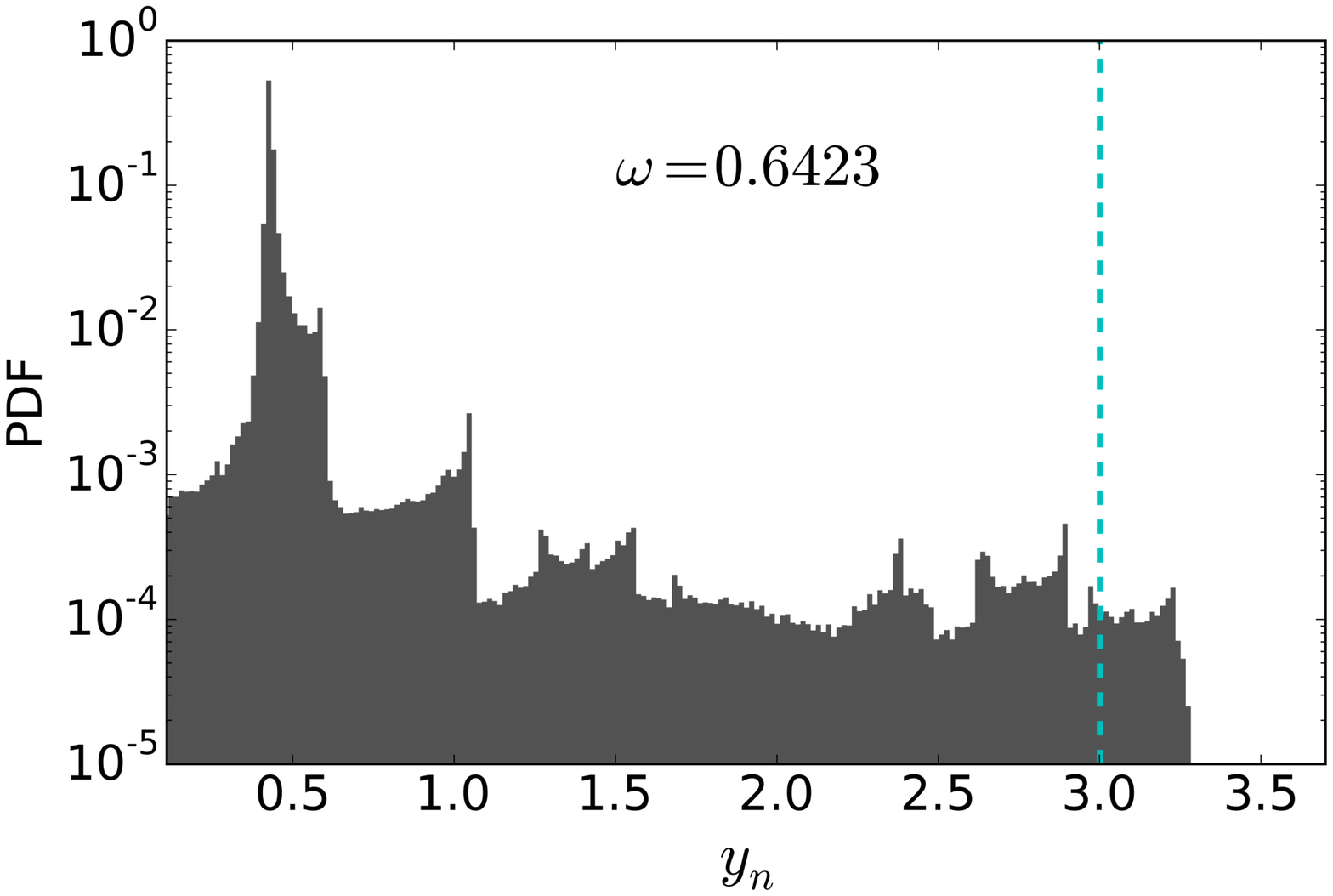}
	\caption{\label{fig:epsart} (color online) PDF of $y_{n}$  for three different $\omega$. Upper panel is for $\omega = 0.735$ where the attractor is chaotic and bounded in a small region in phase space.  Middle panel corresponds to  $\omega = 0.7315$ where extreme events are generated by interior crisis. In lower panel at $\omega = 0.6423$, extreme events occur through PM intermittency from a periodic state. Vertical dashed lines in cyan denote the corresponding $H_{S} (=\mu+8\sigma$) mark.}
\end{figure}

\begin{figure}[]
	\includegraphics[width=7.5cm, height=5cm]{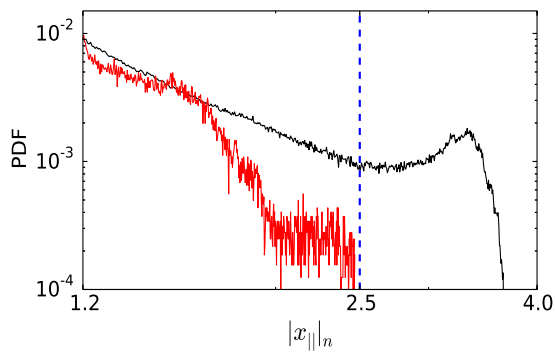}
	\caption{\label{fig:epsart} (color online) PDF of events during extreme events (black) and immediately before  extreme events (red). Distribution  for $k_{1} = k_{2} = - 0.0686$ (red color), when quasiperiodicity breaks down to chaos but no extreme events occur. Blue vertical line denotes the significant height($H_{S}$). For $k_{1} = k_{2} = - 0.07$,  the distribution (black line) shows dragon-king-like extreme events.}
\end{figure}

\begin{figure*}[hbt!]
	\hspace{-80pt}
	\begin{subfigure}[b]{0.5\columnwidth}
		\centering
		\includegraphics[width = 6.25cm,height = 4.75cm]{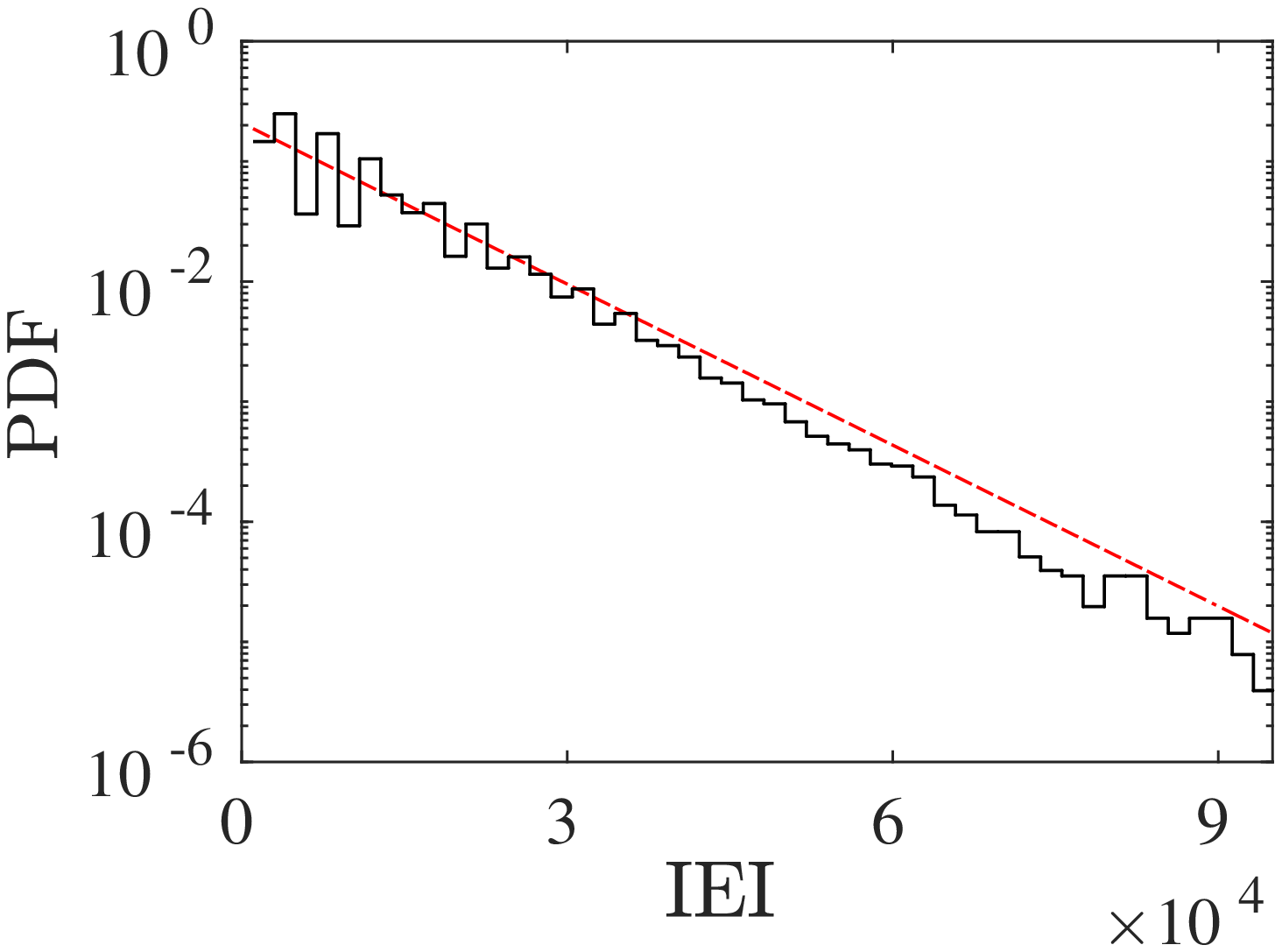}
		\label{}	
	\end{subfigure}
	\hspace{50pt}
	\begin{subfigure}[b]{0.5\columnwidth}
		\centering
		\includegraphics[width = 6.25cm,height = 4.75cm]{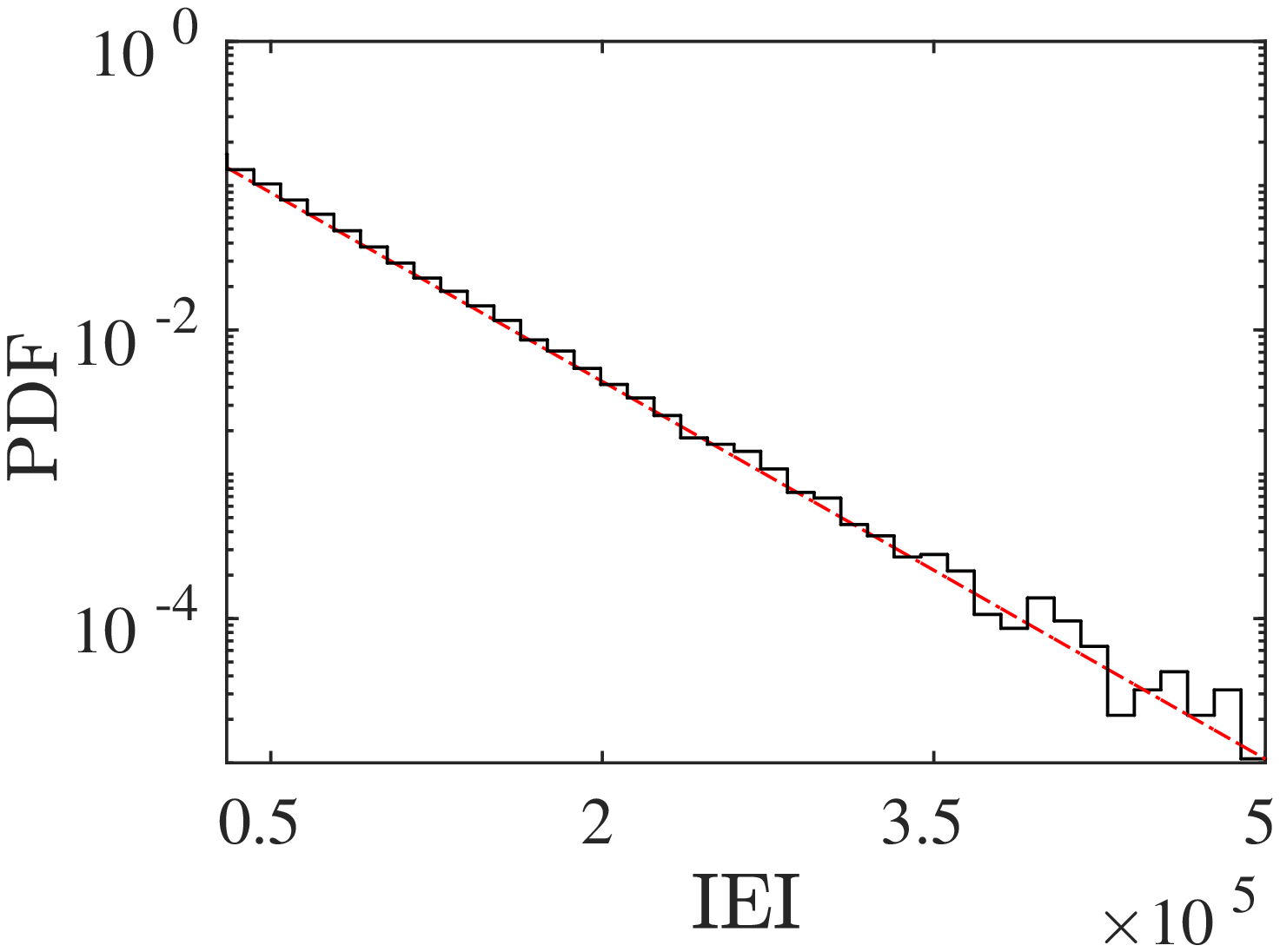}
		\label{}	
	\end{subfigure}\\
	\hspace{0pt}
	\vspace{50pt}
	\hspace{-70pt}
	\begin{subfigure}[b]{0.5\columnwidth}
		\centering
		\includegraphics[width = 6cm, height = 4.5cm]{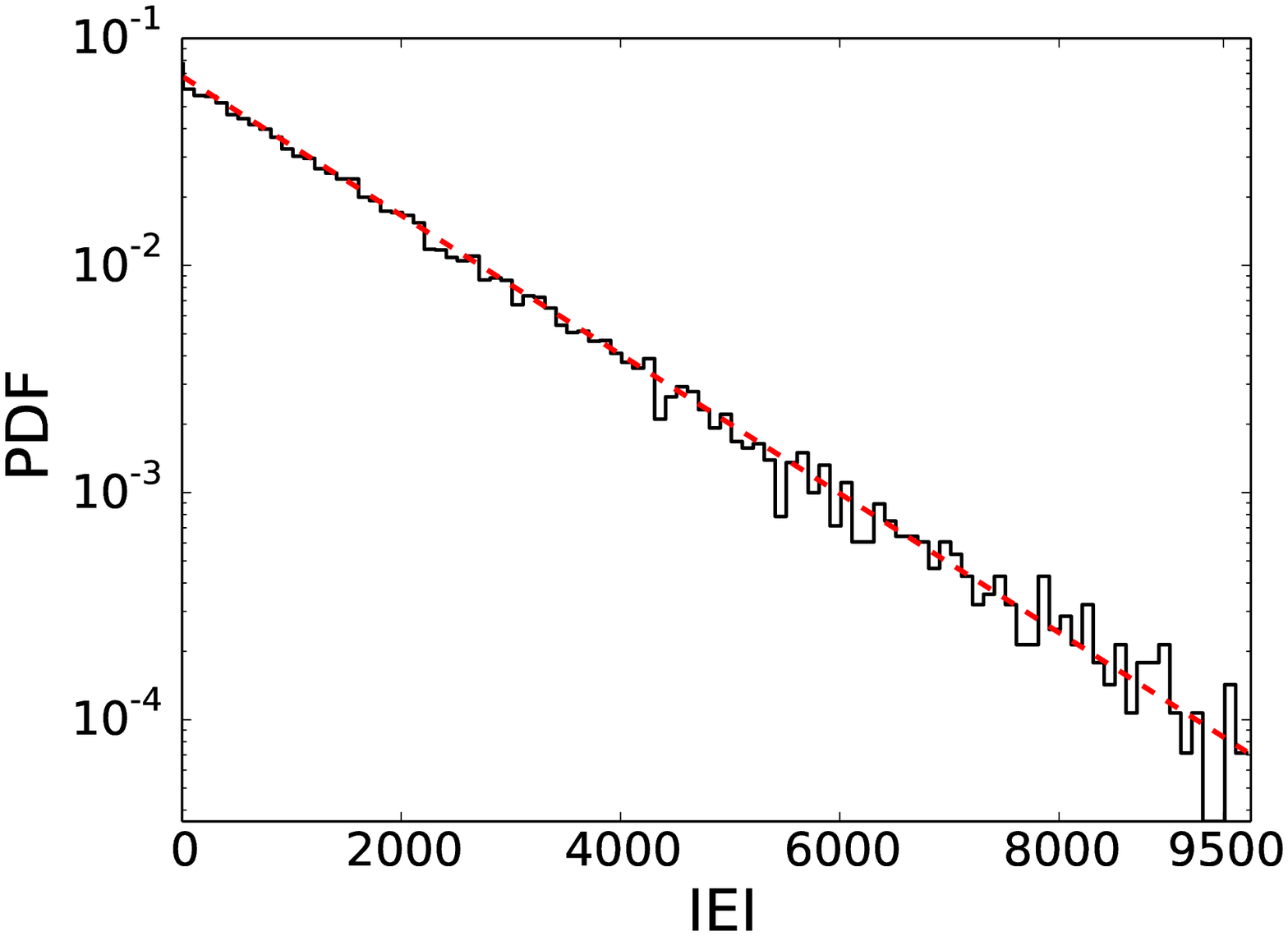}
		\label{}	
	\end{subfigure}
	\hspace{50pt}
	\begin{subfigure}[b]{0.5\columnwidth}
		\centering
		\includegraphics[width =6cm,height = 4.5cm]{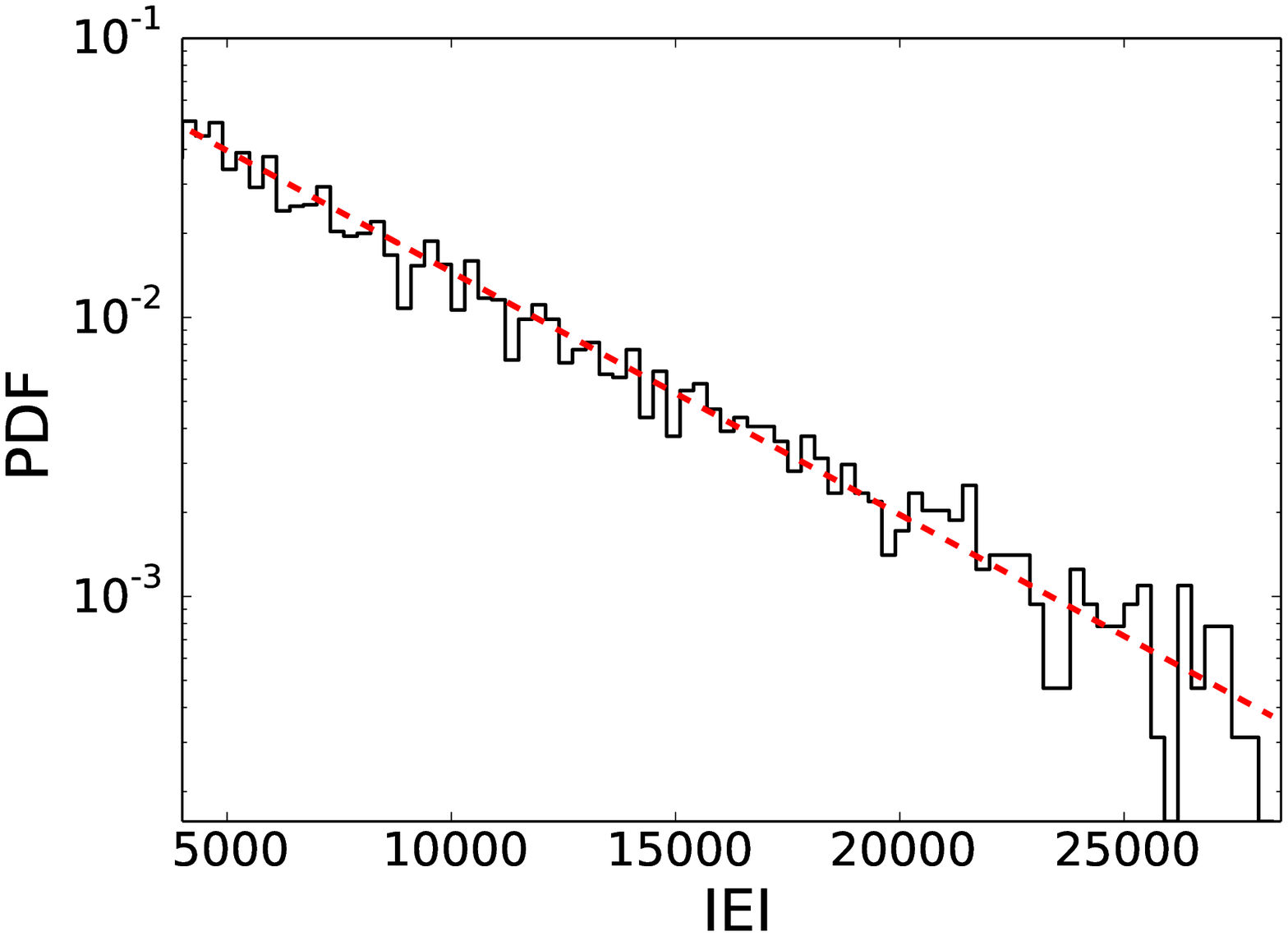}
		\label{}	
	\end{subfigure}
	\vspace{-50pt}
	\caption{(Color online) PDF of inter-event-interval (IEI) of extreme events.  Upper panels represent the forced Li\'enard system: left panel for interior crisis-induced extreme events ($a=0.2084, b=1.029 \times 10^{-4}$), right panel for PM intermittency-induced intermittency ($a=0.2437, b=2.008\times 10^{-5}$). Lower panels represent the coupled neuron model: left panel for PM intermittency-induced extreme events ($a = 0.0681,  b = 7.052\times10^{-4}$) while the right panel is for the quasiperiodic route ($a = 0.1354,  b = 1.947\times10^{-4}$). Red lines indicate linear fit of PDF vs. IEI as defined by  $P(x)=ae^{-bx}$ in a log-normal scale. Number of iteration are taken (after removing long transients) as $5\times 10^9$ for all the cases.   }
	\label{FIG.5}
\end{figure*}
	
The PDF is bounded in the chaotic regime (upper panel, Fig.~6 upper panel, $H_S$ line is marked by vertical dashed line), but it spreads into long-tail when extreme events set in via period-doubling cascades followed by interior crisis (Fig.~6, middle panel). Similarly, we find a long-tail distribution of events when extreme events set in via PM-intermittency route (lower panel, Fig.~6) and even rarer. Interestingly, in the coupled neuron example, we notice a different PDF of events as shown in Fig.~7 for the quasiperiodic route to extreme events. In the bounded chaotic case ($k_1=k_2=-0.686$) shown in Fig.5 (third row from top), the PDF remains bounded (red curve), coincidentally,  within the limit of $H_S (=\mu+6\sigma$) line (vertical dashed line). We note also that the PDF is characterized by two slopes. Once the extreme events set in for increasing inhibition, we notice that the PDF extends to larger event sizes beyond  $H_S$ line. It follows a power-law as long as the events' heights are within the range of the significant height $H_S$ (vertical blue line) and larger events are outliers to the power-law showing signatures of a dragon-king-like characteristic \cite{ott, mishra}. We are yet to understand this transition from a two slope behavior to a power-law behavior (checked using Kolmogorov-Smirnov fitting) and the origin of a dragon-king, which is our focus for a future study. We check that the dragon-king-like behavior is also present in the case of extreme events occurring via PM intermittency in coupled neurons (not shown here). It is to be noted  here that our estimated $H_S (=\mu+6\sigma$) vertical line, for the coupled neurons, drops exactly at a $|x_{||}|$ value above which a transition from a chaotic state to extreme events is seen in Fig.~7. 
Our choice of 6 times $\sigma$ to estimate the threshold $H_S$ line thus makes a correct prediction of the  transition point to the emergence of extreme events. 
The estimate on $H_S$ line is usually made as ($H_S=\mu+d\sigma$) where $d$ is chosen from a variable range of ($4$ to $8$), for different systems, in literature. A confirmation of this choice is always made when we look at  the PDF of event heights and check whether  $H_S$ line really counts the rare events only as confirmed by our examples. PDFs of inter-event-intervals (IEI) of extreme events for all three routes are elaborated for both  systems and depicted in Fig.~8, showing Poisson distributions and confirming the uncorrelated nature of extreme events. Red lines indicate a linear fit of the exponential decay, (P$(x)=ae^{bx}$) of PDF with IEI as shown in a log-normal scale. The parameter values (a, b) are given in the figure caption.

\section{Conclusion} We have discussed three different dynamical processes that trigger a sudden transition to extreme events from a nominal state in response to changes of a parameter. The studied transitions occur as a result of an interior crisis, PM intermittency and the breakdown of quasiperiodicity. 
To illustrate those transitions we employ two different example systems each of them possessing two different routes at different range of parameters: a forced Li\'enard system and a coupled HR neuron model with synaptic coupling. 
Using bifurcation diagrams, temporal dynamics, Poincar\'e surface of section plots as well as return maps, we have shown that the dynamics of the model systems remained bounded to a state of nominal amplitude, but could suddenly change to events with large amplitude when a system parameter reaches a critical value. This large excursion of the trajectory is occasional, i.e. rare, and the trajectories return to the nominal state after a short duration, but  these excursions are recurrent. We have only considered these events as extremes, which qualify according to a statistical definition. Here we have found for all model systems considered here that the usual $99^{th}$ percentile definition used in the statistical literature is rather conservative and therefore we define  significant height $H_S$ as a threshold of event size above which events are only considered as extremes. 
\par Additionally, we have estimated the PDF of events before and after the transition, from a nominal phase to the onset of extreme events. The PDFs of the size of events have confirmed the long-tail behavior establishing the fact that the large events are really rare for all the routes. Exceptionally, in the coupled neuron model, PDF shows a dragon-king like behavior that also exhibits the long-tail property, but additionally follows a power-law within the limit of the bounded size of events; extreme events are outliers. 
PDFs of inter-event-interval, in our example cases,  all follow a Poisson distribution confirming that the extreme events discussed here are uncorrelated.
\par 
 From the findings of previous works \cite{karnatak, cristina, bonatto}  
in different model systems, we searched for the  most common dynamical instabilities that may lead to extreme events. Our findings indicated, at least, but of course not exhaustive, three general types of instabilities,  that lead to extreme events in response to parameter variation of the system.  Examples of dynamical systems with crisis-induced extreme events are abundant, in  literature. On the other hand, there are many examples of intermittent large size spiking or bursting via PM intermittency, but they have never been studied from the angle of extreme events, except in an experimental study in recent time \cite{sujith}. We demonstrated both the routes leading to extreme events with our two supportive examples \cite{kingston, mishra}. The quasiperiodic route to extreme events is less known, on which we find a distinct example in the coupled neuron model under inhibitory synaptic interactions. It is noteworthy that our example systems individually do not reveal all the three routes to extreme events. The Li\'enard system has no quasiperiodic origin of extreme events while the coupled HR neuron model does not show period-doubling route to chaos followed by the interior crisis-induced extreme events, at least, for our choices of system parameters. This reveals the system dependence of the presence of a particular route to the origin of extreme events, where the intrinsic dynamical traits of a system must be playing a dominant role.
\\\\
{\bf AUTHOR'S CONTRIBUTIONS}\\
All the authors contibuted equally to this work.
\\\\
{\bf ACKNOWLEDGEMENTS}\\
T.K. and L.K. have been supported by the National Science Centre, Poland, OPUS Programme Project No. 2018/29/B/STB/00457. S.K.D. is supported by the Division of Dynamics, Lodz University of Technology, Poland. C.H. is supported by INSPIRE-Faculty grant (code: IFA17-PH193). AM is supported by the CSIR(India). U.F. has been supported by Volkswagen Foundation (Grant No. 88459). 
\\\\
{\bf DATA AVAIALABILITY STATEMENTS}\\
The data that supports the findings are all generated by computer simulations by the authors. All data are available from from the corresponding authors upon reasonable request.
\begin{appendices}
	\section{Appendix: Model description}
	\subsection*{Li\'{e}nard system}
	
	The Li\'enard system represents a class of second order differential equations. We consider one such model  whose description is presented  in Sec.II in the main text, with a specific set of parameters. 
	\begin{figure}[hbt!]
		\includegraphics[height=6cm, width=7cm]{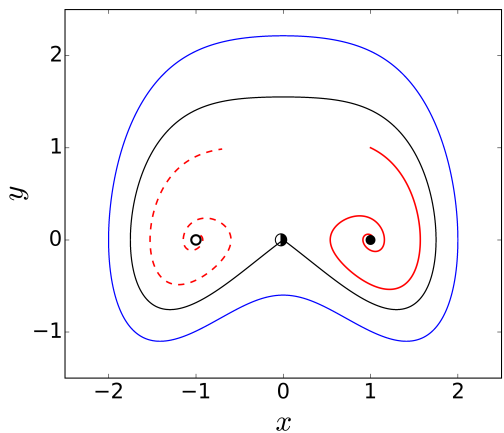} 
		\caption{(color online) Phase portrait  of autonomous Li\'enard system. Black line denotes the homoclinic orbit (HO)  connecting the saddle origin, denoted by half filled circle. Open and solid black circles  represent unstable and stable focus at $(-1,0)$ and $(1,0)$, respectively. Any choice of initial condition outside the HO creates neutrally stable orbit like the one in blue line. }
		\label{FIG.9}
	\end{figure}
 
	For a parametric condition, $\alpha>0$, $\beta>0$ and $\gamma<0$, the autonomous system ($A=0$) shows \cite{kingston, mishra1} a dual character  of dissipative and conservative systems in state space. We  select the parameters as given here as $\alpha~\mathrm{= 0.45}$, $\beta~\mathrm{= 0.50}$ and  $\gamma~\mathrm{= -0.50}$ that satisfy the conditions to maintain the dual character. The region of dissipative dynamics is then confined to an area bounded by a homoclinic orbit (HO)  connected to the saddle origin (black line, Fig.~9). Furthermore, there exist two additional equilibrium points, a stable focus $(1,0)$ (solid black circle, Fig.~9), and a saddle focus $(-1,0)$ (open black circle, Fig.~9) inside the HO. 
	Outside the HO, the system maintains the conservative character of the system, possesses infinitely many coexisting large periodic orbits (an exemplary large orbit shown in blue lines), which are all neutrally stable (for details see refs.\cite{kingston, mishra1}). For any choice of initial conditions inside the HO, the trajectory always converges to the stable focus.
	\par When a periodic forcing ($A\neq0$) is applied to the system, 
	each of the neutrally stable orbits outside the HO turns into a quasiperiodic orbit lying on top of each other in many layers that coexist with a periodic orbit. The periodic orbit emerges when $(1,0)$ becomes unstable as shown in Fig.~3 (upper left panel). The periodic orbit eventually becomes chaotic via period-doubling with decreasing $\omega$ followed by extreme events and periodicity for further decrease in $\omega$ as shown in the bifurcation diagram in Fig.~1. However, the saddle point at origin $(0,0)$ becomes a saddle orbit \cite{guckenheimer}. 
	\subsection*{Hindmarsh-Rose model}
	The Hindmarsh-Rose model \cite{HR} describes the spiking-bursting behaviors of neurons. It is described as a system of three nonlinear ordinary differential equations
	\begin{gather*}
		\dot{x} = y + bx^{2} - ax^{3} - z + I\\
		\dot{y} = c - dx^{2} - y\\
		\dot{z} = r[s(x - x_{R}) - z],
	\end{gather*}
	where $x$ denotes the membrane potential, $y$ and $z$ variable denotes transport of ions through the ion channels. The transport of sodium and potassium ($Na^+$ or $K^+$) ions is made through fast ion channels and its rate is measured by $y$, which is called the spiking variable. The transport of slow calcium ions ($Ca^{++}$), which is taken into account in the evolution of the $z$-variable,  controls the fast spiking in $x$- and $y$-variable and thereby originates bursting oscillations. We choose the parameters as, $a = 1, b = 3, c = 1, d = 5, x_{R} = -1.6, r = 0.01, s = 5, I = 4$ when a single neuron shows periodic bursting with three spikes in a burst as shown in Fig.~10. Here, we consider two identical HR systems with three periodic bursting and couple them with chemical synaptic coupling and observe the  dynamics with excitatory and inhibitory synapses, separately.
	\begin{figure}
		\includegraphics[height=6cm, width=9cm]{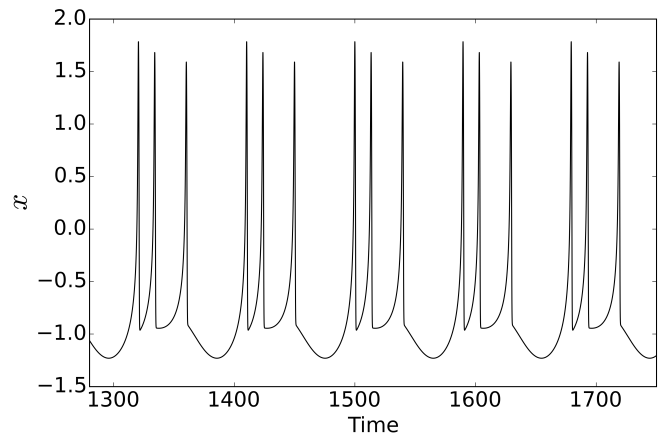} 
		\caption{(color online) Time series of the $x$ variable for a single Hindmarsh-Rose oscillator. The time series show periodic bursting behavior with three spikes in a burst.}
		\label{FIG.10}
	\end{figure}
	\section{Appendix: Computational Methods}
	This work is based on numerical experiments using Python. We use the fourth order Runge-Kutta method with a fixed time step of $0.01$. We also check the consistency of our results	using an adaptive integrator. The initial conditions are chosen as random. For generating time series and plotting Poincar\'{e} surface we consider $2\times 10^{6}$ data points after discarding the transients ($10^6$ iterations). For plotting probability distribution function (PDF) of the event size and inter event interval (IEI), the simulation is run for $5\times 10^{9}$ time steps. Within this time length, we detect local maxima  and their corresponding time by peak detecting algorithm from the time series and consider the local maxima as events. For handling groups or bursts of EE, we consider the highest peak within a single burst and discard all other maxima as described in \cite{ott}. Now we have event series with corresponding time. From this series, we plot the PDFs of event size and also the PDFs of the IEI. The significant height($H_{s}$) is also calculated from this event series using a definition, $H_{s} = \mu + d\sigma$ ($d$ is a constant), where $\mu$ is the mean of the event series and $\sigma$ is the corresponding standard deviation. This estimate is system dependent where $d$ is arbitrarily chosen\cite{kharif, muller} in a range of 4-8; for the Li\'enard system $d$=8 and for the coupled HR model $d$=6 is used consistently for all the simulations for each system. Any event which crosses this $H_{s}$ line, is considered as an extreme event. We also check that within the length of our simulation the PDFs reach statistical convergence and become stationary, i.e, if we add more data points the shapes of the PDFs do not change significantly.
\end{appendices}

\end{document}